\documentclass[amsmath,trackchanges, twocolumn, twocolappendix]{aastex702}

\usepackage{newtxtext,newtxmath}
\usepackage{diagbox}
\usepackage{hyperref}
\usepackage{orcidlink}
\usepackage{graphicx}	% Including figure files
\usepackage{amsmath}	% Advanced maths commands
\usepackage{mathtools}

\shorttitle{Observational Signatures of Giant Impacts}
\shortauthors{L. Taylor et al.}
\begin{document}

\title{Blinding Lights: Simulation-Based Evidence for Terrestrial Giant Impacts as Astronomically Observable Events}

\author[orcid=0009-0007-3737-9674,gname=Lucy,sname=Taylor]{Lucy Taylor}
\affiliation{School of Physics, H.H. Wills Physics Laboratory, University of Bristol, Bristol BS8 1TL, UK}
\email[show]{lucy.taylor@bristol.ac.uk}  

\author[orcid=0000-0001-5365-9616]{Simon J. Lock} 
\affiliation{School of Earth Sciences, University of Bristol, Bristol, BS8 1RJ, UK}
\email{s.lock@bristol.ac.uk}

\author[orcid=0000-0001-5065-4625]{Philip J. Carter}
\affiliation{School of Physics, H.H. Wills Physics Laboratory, University of Bristol, Bristol BS8 1TL, UK}
\email{p.carter@bristol.ac.uk}

\author[orcid=0000-0003-4813-7922]{Zo\"e M. Leinhardt}
\affiliation{School of Physics, H.H. Wills Physics Laboratory, University of Bristol, Bristol BS8 1TL, UK}
\email{Zoe.Leinhardt@bristol.ac.uk}

\begin{abstract}
Giant impacts (collisions between planet-sized bodies) are violent events that fundamentally alter the properties of planets and the structure of planetary systems. Observations of these events provide important constraints on the role of giant impacts in the formation of planetary systems. However, relating such observations to the specifics of the impact responsible is challenging, since numerical simulations struggle to resolve the expected extreme low densities of observable material at the edge of ongoing impact structures. In particular, lower limits on density are often enforced in SPH codes for computational efficiency. Here, we present a method of post-processing output from smoothed particle hydrodynamics (SPH) simulations of giant impacts, that emulates realistic physical properties in multi-phase and multi-material collisions, in order to better understand these low-density outer regions. We apply our novel post-processing pipeline to determine the direct thermal emission and potential observational signatures of giant impacts, and obtain consistent photic surface geometries for an example impact simulated for a range of resolutions and code parameters. We find that during the first days after this terrestrial giant impact event, the emission from hot, unbound ejecta can be comparable in temperature and luminosity to that of an M-Dwarf star, and as such can be astronomically observable. Our work begins to bridge the gap between observations and theoretical models, and could allow high cadence, large-sky surveys -- such as that to be carried out by the Vera C. Rubin Observatory -- to place constraints on the frequency and properties of giant impacts in our local neighbourhood. 
\end{abstract}



\section{Introduction} 
Giant impacts (collisions between planet-sized bodies) are the most violent events in planetary formation. In collisions between planets of a few Mars masses or greater, the kinetic energy of the colliding planets can be converted to thermal energy at the same rate as power is emitted from the Sun for several hours after the impact event \citep{Carter+2020}. Giant impacts are so energetic that the gravitationally bound post-impact remnant can be substantially vaporised and rapidly rotating, reaching incredibly high temperatures and low densities \citep{LockStewart17,caracas_stewart_2023,postema+2026}. Such objects are many tens, if not hundreds, of times larger than the original colliding bodies \citep{LockStewart17,lock2018_moon}. In addition, large amounts of unbound debris are produced, substantial fractions of which are initially vaporised \citep{asphaug+2006, leinhardt_stewart_2012, Genda+2015, carter+2026}. This debris expands outwards from the impact point and produces structures that may eventually span the entire host system \citep{jackson_wyatt_2012, watt+2023}. The epoch of giant impacts is usually considered to be planet formation in its final stages, yet these events can be so extreme as to significantly alter the structure, composition and orbits of forming planets. As such, giant impacts can completely reshape planetary systems.

There is strong evidence that giant impacts have occurred in our own solar system – for example, Uranus’ atypically high obliquity \citep{safronov66,slatterly1992, Kegerreis+2018, Rogoszinski_2021}, Mercury’s thin mantle \citep{Benz88_Mercury}, as well as the existence, size, composition and eccentricity of Earth’s moon \citep{moon_hartmann_davis_75,moon_cameron_ward_76,simon_moon_review_paper_2020}. These features are all hypothesised to be the result of one or multiple giant impacts. It is very reasonable to think that giant impacts likely take place in other solar systems too \citep{Raymond_2014, Schlichting_2018, bonomo+2019, naponiello+2023}, but it would be incredibly remiss to simply extrapolate the giant impact record of our own solar system to general planet formation pathways. Observations of exoplanetary systems have demonstrated that the orbital architecture and constituent planets of our solar system are probably unusual \citep[e.g.][]{Winn2014,Fulton2017,valencia_diversity_2025}. The rapidly increasing database of known exoplanets has revealed several common classes of planet that are absent from our solar system, for example super-Earths and mini-Neptunes, whose masses populate the observed gap between the masses of Earth and Neptune. Furthermore, many exosystems are much more compact in orbital semi-major axis than our own \citep{Raymond_2018-20}.

Although observational biases in exoplanet detection techniques make it difficult to ascertain the exact frequency of different orbital architectures, these discoveries have prompted the re-examination and proliferation of planetary formation models in order to explain the formation of systems both like and unlike our own. Crucially, different theoretical planet formation scenarios predict different giant impact occurrence rates, with some models considering giant impacts to be dynamical inevitabilities that are the main avenue of planetary growth \citep{Raymond2009,Walsh2011,Quintana+2016,clement_rethinking_2022}, whilst others posit that a giant impact phase is not always necessary to form terrestrial or giant planets \citep{Lambrechts+2012, bitsch+2019}, with a spectrum of models in between \citep{inamdar_schlichting_2015, Izidoro+2017, shibata+2025}. In other words, we do not yet know whether giant impacts are a key feature of the recipe for planet formation, or simply optional decoration.

Recent, and anticipated future, observations of the aftermath of impacts in exosystems have the potential to constrain the frequency and role of giant impacts in planet formation, as well as elucidating general planet formation pathways. Evidence of impacts -- giant or otherwise -- is often inferred from observations of circumstellar dust and gas exhibiting thermal, and sometimes compositional, properties incompatible with that of a quiescent system \citep[e.g.][]{meng+2012,meng+2015, Su2019, Su2022, Schneiderman+2021, Chen+2024,tzanidakis_gaia-gic-1_2026}. Such signals are generally attributed to material that has been ejected from planetary atmospheres and interiors during impact events, and then passively heated by the system's star. However, it has recently been suggested that the observed infrared brightening of star ASASSN-21qj was due to direct emission from a hot post-impact remnant whose bound mass spanned hundreds of Earth radii, rather than passively heated dust several hundred days later. Furthermore, a complex optical dimming (at times, a near-total occultation) was interpreted as the subsequent transit of impact debris \citep{KenworthyLock23}. Although work has been done to understand the long-term evolution of (unbound) debris from giant impacts \citep{jackson_wyatt_2012,watt+2023,carter+2026}, we currently lack the tools required to validate and interpret direct observations of the early stages of giant impacts or their (bound) remnants. The impact around ASASSN-21qj is thought to have involved volatile-rich super-Earths or mini-Neptunes, but detection of the direct emission from a post-impact body opens the possibility that direct detection of the hot clouds of vapour produced by giant impacts between a range of different bodies could be detected by the next generation of high-cadence, large-sky surveys -- such as that to be carried out at the Vera C. Rubin Observatory \citep{rubin_mission}. Fortuitous transits of extended post-impact bodies across their host stars might provide yet another avenue for their detection.

\begin{figure*}
    \centering
    \includegraphics[width=\textwidth]{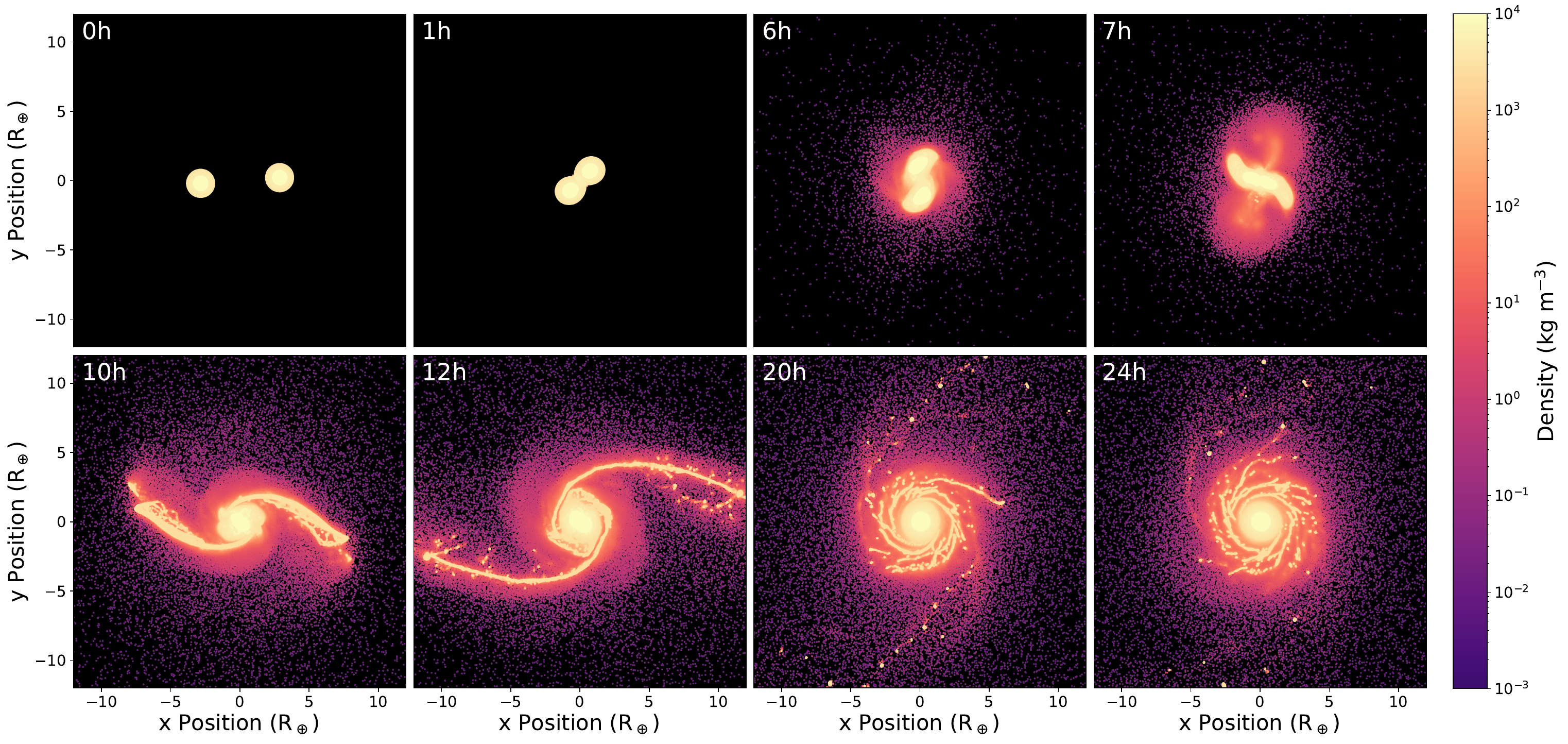}
    \caption{Eight snapshots from the first 24 hours of a planetary collision (taking cut-throughs at the midplane of the collision) simulated using \textsc{swift}. Each particle is illustrated as a single point with a colour determined by their central density. The point of impact occurs at 1h in simulation time. The target and impactor are identical ``proto-Earth-like'' 2-layered (iron core, forsterite mantle) planetary objects $\sim1 \rm R_{\oplus}$ in radius and $\sim0.9 \rm M_{\oplus}$ in mass. Each colliding planet is made up of $\sim 10^6$ SPH particles. The impact speed is the mutual escape speed, and the impact parameter is $0.7$. The maximum smoothing length that a particle can have ($h_{\rm max}$, see Section \ref{sect:sph_intro}) is set to 1~$\rm R_{\oplus}$.}
    \label{fig:giant_impact_example}
\end{figure*}

In this paper, we present a method for accurately predicting the observational signatures of giant impacts from smoothed particle hydrodynamics (SPH) simulations, SPH being the primary tool used in the study of giant impacts (see Figure~\ref{fig:giant_impact_example}) across an extensive range of contexts and applications \citep[e.g.][]{Canup2001, lock2018_moon, Reinhardt_2020, Nakajima_2021, dou+2024, Bussmann_2025, roche+2025, Sandnes_2025}. Making the connection between models and observations is inherently difficult, since a simulation-based study of giant impact observables requires accurate tracking of the properties of material in the outermost regions of ongoing impact structures. This material is expected to have an extremely low density; we show in Section \ref{subsect:data_processing} the photosphere can be defined by material of densities of $\mathcal{O}(10^{-8})$~kg~m$^{-3}$ and even below. Even with the use of state-of-the-art SPH codes, this presents a significant challenge, particularly as many SPH codes make necessary compromises between computational efficiency and the lowest density of material that is  accurately tracked (we discuss this in depth in Section~\ref{sect:sph_intro}). To overcome this obstacle, we introduce a post-processing step to emulate more realistic density ranges. This forms the basis of a longer post-processing pipeline which untangles the multiphase, multi-material thermal properties at any location in an ongoing impact system, extracts the outer edge of the optically thick -- and hence emitting -- regions, and so quantifies the emission and/or absorption that could make a given impact observable. 

In Section~\ref{sect:sph_intro}, we review the key principles of SPH codes and the problems that can be incurred by common simplifications implemented for sake of numerical efficiency. We then outline our post-processing method used to robustly extract realistic giant impact observables (Section~\ref{sect:methods}) and present the application of our new pipeline to an example giant impact scenario between two equal-mass rocky planets (Section~\ref{sect:analysis}). We analyse the consistency of the results yielded by our pipeline -- both in terms of thermal properties and photic surface geometries -- for SPH simulations of an example impact event using a wide range of numerical resolutions and code parameters, and explore the evolution of impact observables throughout the course of the impact. In Section \ref{sect:discussion}, we discuss the implications of our results for verifying and interpreting giant impact observations and the current limitations imposed on our method and current SPH simulations. We conclude in Section \ref{sect:conclusions}. Additional technical methodology, tables and figures are given in appendices. Supporting information containing scripts to reproduce the simulations, analysis, and figures from this work will be made available through GitHub and Zenodo upon acceptance.

\section{Key Principles of SPH and Limitations in Implementation}
\label{sect:sph_intro}
In this section, we discuss the key concepts of SPH that motivate our post-processing pipeline. Since this work uses simulations conducted with the SPH code \textsc{swift} \citep{2018swift}, we discuss relevant features in the context of the SPH implementation used in \textsc{swift}. However, similar principles apply to many other SPH codes, and it would be straightforward to apply our pipeline in other cases. 

An example of output from a \textsc{swift}-simulated giant impact is shown in Figure \ref{fig:giant_impact_example}. In SPH simulations of giant impacts, the total mass of each colliding planet is divided into parcels of mass known as particles. Particles are usually chosen to be of approximately equal mass, since numerical issues can arise when unequal mass particles interact. In this work, we will use $N_{\rm tot}$ to refer to the total number of particles in a simulation, and $N$ to refer to the number of particles in the target planet (i.e. the largest planet)  involved in a giant impact.

SPH particles are represented by spheres of radius $H$ (the kernel support radius, which is finite in most computational applications), but with their mass not uniformly distributed within that volume. Instead, a radial mass distribution is determined by a smoothing kernel $W(\mathbf{x}, h)$, which is normalised over the particle volume (see Appendix~\ref{appendix:sim_data_table}). Here, $\mathbf{x}$ is a vector measured from the central location of the SPH particle. $h$ is the so-called smoothing length of the particle, and is related to the kernel support radius by a constant, such that $H = \gamma h$, with $\gamma$ dependent on the choice of kernel function. Furthermore, most SPH kernel functions can be rescaled so that they can be thought of as operating on vectors in the unit sphere \citep{dehnen_aly_2012}:

\begin{equation}
    W(\mathbf{x}, h) = H^{-3} \ w(|\mathbf{x}|/H) \; ,
\end{equation}
where $w$ is dimensionless. In this work, our simulations use a cubic spline kernel, $b_4$, which is normalised such that $w = 16 \pi^{-1} \ b_4$ (see Appendix~\ref{appendix:sim_data_table}). For more examples of kernel functions and their corresponding formulae and normalisation constants, see Table 1 in \cite{dehnen_aly_2012}.

The density, $\rho_i$, at the centre of an SPH particle $i$, with smoothing length $h_i$ at location $\mathbf{x}_i$ is approximated by
\begin{equation}
    \label{eqn:sph_rho}
    \rho_i = \sum_{j} m_{j} W(\mathbf{x}_i - \mathbf{x}_j, h_i) \; ,
\end{equation}
with the index $j$ summing over all SPH particles ($i$ inclusive), whose individual masses are $m_j$. Note that only particles whose central location $\mathbf{x}_j$ is within the kernel support radius of particle $i$ will have a non-zero contribution to the density $\rho_i$.

SPH can interpolate a range of material properties besides density. As demonstrated in \cite{Price_2012}, to approximate an arbitrary scalar function $A$ at a location $\mathbf{x}$, we can interpolate between all SPH particles (indexed by $j$) as follows:
\begin{equation}
    \label{eqn:sph_interpolant}
    A(\mathbf{x}) = \sum_{j} m_{j} \frac{A_j}{\rho_j} W(\mathbf{x} - \mathbf{x}_j, h) = \sum_{j} V_{j} A_j W(\mathbf{x} - \mathbf{x}_j, h) \;,
\end{equation}
where for all $j$, $A_j$ is the property $A$ evaluated at the centre of the SPH particle $j$, and $V_j$ is an effective volume of particle $j$. Note that smoothing length $h$ is a property of the position being considered and does not change with $j$, and that the kernel $W$ is effectively centred at $\mathbf{x}$, since weightings are being determined by distances of interpolation points from $\mathbf{x}$. Interpolants of scalar properties evaluated at the central locations of SPH particles use the kernel defined by their associated particle smoothing lengths. Indeed, if we consider the case that $A =\rho_i$ for a particle centrally located at position $\mathbf{x}_i$ with smoothing length $h_i$, Equation~\ref{eqn:sph_interpolant} reduces to Equation~\ref{eqn:sph_rho}.

Particle locations and smoothing lengths are updated throughout a simulation: locations according to inertia, and gravitational and pressure forces, and smoothing lengths according to the corresponding change in the number density, $n$, of SPH particles. Generally speaking, in areas of lower number density, an SPH particle will have a larger smoothing length, and hence a larger kernel support radius (the exception to this generalisation is discussed later in this section). Conceptually, increasing the support kernel radius increases the volume over which the (fixed) particle mass is distributed.

At a given simulation time, for a given set of particle locations, $\{ \mathbf{x}_j \ |\ 1 \leq j \leq N_{\rm tot} \}$, most SPH codes, including \textsc{swift}, iteratively solve Equations~\ref{eqn:sph_iterative_h_solve_a}~\&~\ref{eqn:sph_iterative_h_solve_b} to assign each particle $i$ its smoothing length $h_i$ \cite[e.g.,][]{Price_2012}: 

\begin{align}
    \label{eqn:sph_iterative_h_solve_a}
    n_i & = \sum_{j} W(\mathbf{x}_i - \mathbf{x}_j, h_i) \; ,\\
    \label{eqn:sph_iterative_h_solve_b}
    h_i & = \eta\  n_i^{-\frac{1}{3}}
\end{align}
This ensures that $h_i$ is consistent with the number density at $\mathbf{x}_i$, $n_i$. As before, only particles whose location $\mathbf{x}_j$ is within the kernel support radius will have a non-zero contribution to number density. The proportionality constant $\eta$ in Equation~\ref{eqn:sph_iterative_h_solve_b} determines the ``smoothness'' of simulations. In our \textsc{swift} simulations, the masses of individual SPH particles are chosen to be approximately equal, and hence the mean particle spacing, $\big(m_i {\rho_i}^{-1} \big) ^{\frac{1}{3}}$, essentially the cube root of the particle's effective volume, can be  approximated as $n_i^{-\frac{1}{3}}$ (Equations~\ref{eqn:sph_rho}~\&~\ref{eqn:sph_iterative_h_solve_a}). Therefore in \textsc{swift} simulations of giant impacts, the user-specified constant of proportionality $\eta$ is the desired scaling between $h_i$ and the cube root of effective particle volume. Again, it should be emphasised that this is only an appropriate definition of $\eta$ for the case of equal mass SPH particles. Furthermore, the number of neighbours, $N_{\rm neigh}^{i}$, contained within the kernel support radius of a particle $i$ can be approximated as
\begin{equation}
    N_{\rm neigh}^{i} \approx n_i\cdot\frac{4}{3}\pi (\gamma h_i)^3
\end{equation}
i.e., the number density at the centre of the particle multiplied by the particle's volume. Then, using Equation \ref{eqn:sph_iterative_h_solve_b}, we find that
\begin{equation}
    \label{eqn:neighbour_number}
    N_{\rm neigh}^{i} \approx \frac{4}{3} \pi (\gamma \eta)^3.
\end{equation}

\noindent In other words, $N_{\rm neigh}^{i}$ should be approximately constant. For the values used in this study, $\gamma =  1.825742$ and $\eta = 1.2348$, we would have $N_{\rm neigh} \approx 48$. For implementations of SPH that use equal mass particles, estimating $N_{\rm neigh}$ in this way can be a good indicator of how the choice of $\eta$ has affected the smoothness of a simulation. Nonetheless, contrary to what is sometimes stated, having a constant number of neighbours within the kernel support radius is neither enforced by the implementation of SPH used in \textsc{swift}, nor is it consistently correct.

For the sake of computational efficiency, many SPH codes (including \textsc{swift}) do not always assign every particle the value of $h$ that would solve Equations~\ref{eqn:sph_iterative_h_solve_a}~\&~\ref{eqn:sph_iterative_h_solve_b}. For high resolution simulations, calculating properties like Equation~\ref{eqn:sph_rho} would become costly for large smoothing lengths, due to interpolation over a large number of particles. Instead, the user must specify a maximum smoothing length, $h_{\rm max}$, and at no point in the simulation is any particle assigned a smoothing length greater than $h_{\rm max}$. Restricting the smoothing length also restricts the kernel support radius. Imposing an upper limit on smoothing length can hence reduce the number of particles that have a non-zero contribution to interpolated properties, allowing codes to save on computational expense.

Whilst truncating smoothing lengths is often a necessary compromise in the running of SPH simulations, it comes with a cost. When smoothing lengths are capped, some particles become ``isolated'' with effectively no neighbouring particles - i.e., no other particle's central location is within their kernel support radius. In the context of simulations of giant impacts, which involve extensive vaporisation and dispersal of material, smoothing length restrictions can result in a significant portion of particles becoming isolated as a simulation progresses. From Equation~\ref{eqn:sph_rho}, the density of a particle $i$ that has become isolated is given by:
\begin{equation}
    \label{eqn:iso_rho_gen}
    \rho_{i}^{\rm iso} = m_i W(\mathbf{0}, h_i) \; .
\end{equation}

For \textsc{swift} simulations that use the default cubic spline smoothing kernel, $b_4$, the density of an isolated particle is therefore equal to:
\begin{equation}
    \label{eqn:iso_rho}
    \rho_{i}^{\rm iso} = \frac{m_i}{H_i^{3}} w(0) = \frac{m_i}{ H_i^{3}} \ C b_{4}(0) = \frac{ 8 m_i}{ \pi(\gamma h_i)^{3} }
\end{equation} 
where $C = 16/ \pi$ and $b_4(0) = 0.5$ (see Appendix~\ref{appendix:sim_data_table}), and $H_i~=~\gamma h_i$.  If each SPH particle has been assigned approximately equal masses, we can rewrite Equation~\ref{eqn:iso_rho} in terms of total target mass, $M_{\rm tar}$, and target resolution, $N$, and find that the density of an isolated particle is given by:
\begin{equation}
    \label{eqn:iso_rho_approx}
    \rho_{i}^{iso} \approx \frac{8M_{\rm tar}}{\pi N (\gamma h_i)^3} \; .
\end{equation} 

For SPH codes that implement maximum smoothing lengths, Equation~\ref{eqn:iso_rho_approx} is minimised by $h_i = h_{\rm max}$. Therefore, we can estimate the minimum density, $\rho_{\rm min}$, that can be assigned to any SPH particle in a simulation as: 
\begin{equation}
    \label{eqn:rho_min_approx}
    \rho_{\rm min} \approx \frac{8M_{\rm tar}}{\pi N (\gamma h_{\rm max})^3} \;,
\end{equation}
where we have assumed that particles from the target and particles from the impactor have the same mass. In reality, the particle masses in the target and impactor may be marginally different, although this is usually a very minor discrepancy if care has been taken to ensure approximately equal mass particles simulation-wide during initialisation (Section~\ref{subsect:giant_impact_how_to}). As such, we can replace $\frac{M_{\rm tar}}{N}$ in Equation~\ref{eqn:rho_min_approx} with $\frac{M_{\rm tot}}{N_{\rm tot}}$, and find essentially the same value of $\rho_{\rm min}$.

Equation~\ref{eqn:rho_min_approx} shows that by implementing a maximum smoothing length in simulations with equal-mass particles, we concede a ``built-in'' lower limit for SPH particle density before the simulation has even begun, with $\rho_{\rm min}$ often referred to as the density floor. This limit is dependent on the user's choice of $N$, $h_{\rm max}$, and $M_{\rm tar}$.

Although the lower density limit in Equation~\ref{eqn:rho_min_approx} has been derived for a specific smoothing kernel, similar logic can be used to approximate the density floor for any implementation of SPH that relies on equal mass particles and smoothing length caps, including most \textsc{swift} hydrodynamics schemes used in impact simulations. Indeed, this feature is generally analogous to the density floor or ``cut-off density'' in the FDPS SPH code, which is a primary focus of \cite{Hull+2024}, although with a different numerical implementation.

\begin{figure}
    \includegraphics[width=\linewidth]{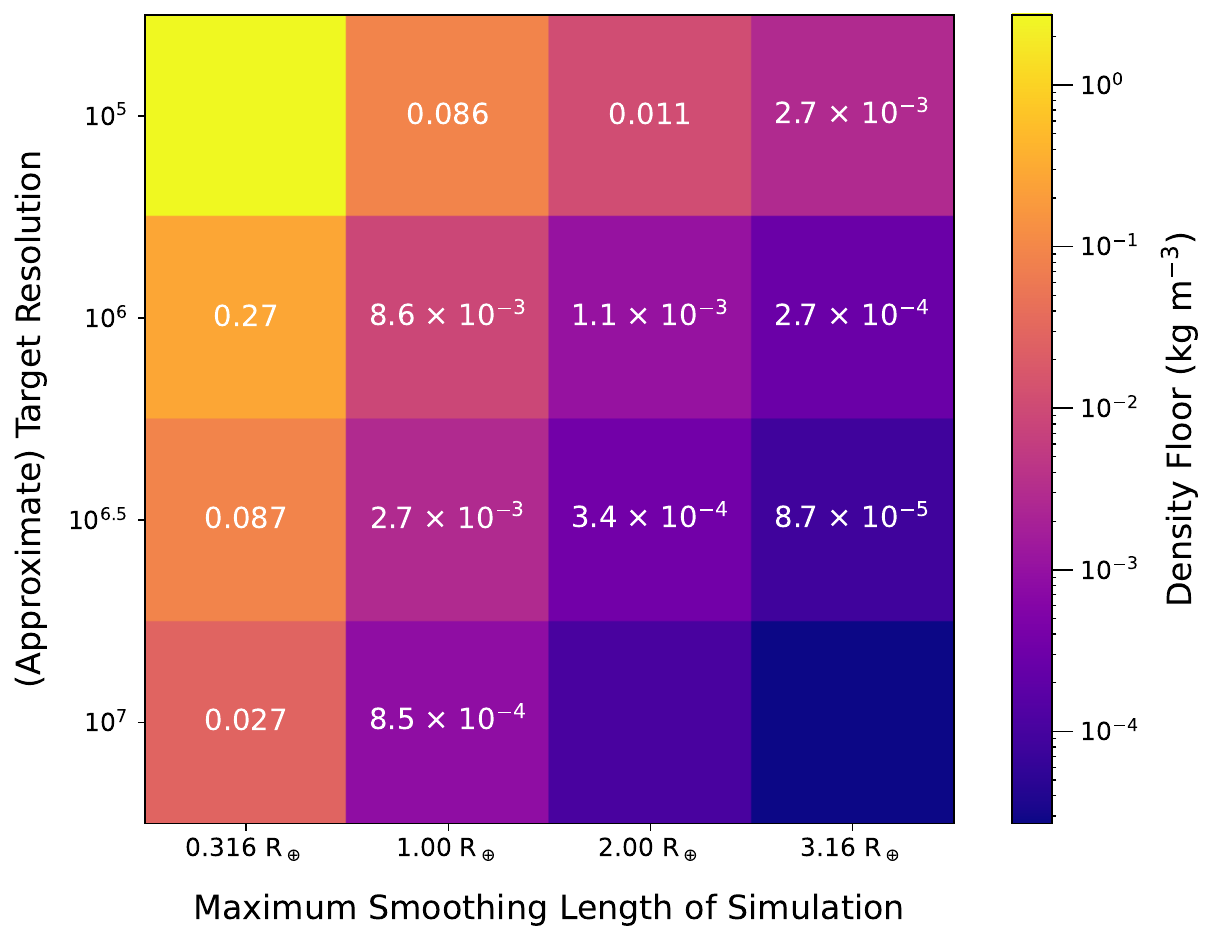}
    \caption{A summary of the target resolutions and maximum smoothing lengths that have been explored as part of this study. Rows are approximate target resolution $N$, and columns correspond to constant maximum smoothing lengths, $h_{\rm{max}}$. The grid is colour-coded by the density floor that each parameter pair corresponds to, and density floor values are specified (to two significant figures) for simulations run as part of this study. We find that the minimum density reached in each simulation is well-approximated by the density floors calculated using Equation \ref{eqn:rho_min_approx}, demonstrating that this limit is essentially hard-coded into the simulation through user-specified parameters.}
    \label{fig:impact_suite}
\end{figure}

As well as some particles storing unrealistically high density values, there are further ramifications of imposing $h_{\rm max}$ for step-wise evolution of SPH particles. Whilst a particle is isolated, it does not experience pressure from nearby particles, and so its motion is only controlled by inertia and gravitational forces. Effectively, the motion of isolated particles is determined in an $N$-body sense only and follows ballistic motion. For isolated particles being held at $h_{\rm max}$ and hence the density floor, no thermodynamic ($P\mathrm{d}V$) work is done on them, and the rate of change of thermal energy is calculated to be 0 \citep[e.g.][]{springel2010_sphreview}. In other words, whilst a particle is isolated and at the density floor, it effectively stops being evolved thermodynamically. Furthermore, whilst non-isolated $h_{\rm max}$ particles are still being influenced by their neighbours, their neighbours have been artificially restricted. As a result, the thermodynamic properties (e.g. density, temperature, pressure) stored by truncated particles, isolated or otherwise, are not representative of the material that would be at those locations if $h$ had been allowed to evolve self-consistently. This limitation, although prevalent in many SPH implementations, is often overlooked, and overcoming it will be a key aspect of our post-processing (Section~\ref{sect:methods}).

\section{Methods}
\label{sect:methods}

In this work we use data from SPH simulations to locate and analyse the photic surfaces of material in ongoing collisions, i.e. the surface at which the optical depth is unity for light entering or leaving the impact structure in a particular direction. Giant impacts between Mars-mass objects or larger can produce significant amounts of silicate vapour \citep[e.g.][]{Davies2017} and hot, extensive objects hundreds of times larger than the colliding bodies in the hours following the collision \citep{LockStewart17}. There are two key challenges in calculating the photic surface from SPH simulations of such impacts:

\begin{enumerate}
    \item[1)] Photic surfaces can be hundreds of planetary radii across, with the transition from optically thick to optically thin often occurring in the outer regions of the post-impact structure where material is of very low density. To capture the dynamics, density structure, and thermodynamics of these outer regions, we therefore require very high resolution simulations.
    \item[2)] The outer regions of the post-impact structure can be a mixture of different materials (e.g. metals and silicates) in multiple phases (e.g. liquid and vapour). This complicates the process of quantifying the material properties required to calculate opacity at relevant locations.
\end{enumerate}

Overcoming these challenges is made more difficult by the fact that, as discussed in Section~\ref{sect:sph_intro}, many SPH codes make higher resolutions more computationally tractable by imposing a maximum smoothing length, and therefore a minimum density. The inclusion of an $h_{\rm max}$ value means that, although a simulation might have particles in a given region, the densities being stored by the particles in that region are not necessarily accurate, nor are the temperatures, pressures, or phases.

In this section, we introduce our pipeline for the post-processing and analysis of simulation data, which overcomes the aforementioned challenges in emulating realistic material properties. We also describe how we use our post-processed data to locate the outermost optically thick surface of an impact structure when viewed from a given angle. We begin by detailing how the giant impacts featured in this study were simulated (Section \ref{subsect:giant_impact_how_to}). We then describe the general theory used to ascertain optical thickness along a given line of sight (Section \ref{subsect:ot_theory}). Combining this optical theory with the limitations of our impact simulations, it becomes clear that unprocessed simulation data precludes the accurate assessment of optical thickness. A post-processing pipeline to circumvent this is presented in Section \ref{subsect:data_processing}, building on the principles discussed in Section \ref{sect:sph_intro}. Furthermore, we develop tools that can interpolate the thermal and material properties needed to determine opacity at any given location in the simulated impact structure. Finally, in Section \ref{subsect:photosphere_emission_method}, we discuss how we calculate the overall thermal properties of the extracted photic surfaces and quantify their emission.

\subsection{Hydrodynamic Giant Impact Simulations}
\label{subsect:giant_impact_how_to}
All simulations of giant impacts presented in this research were simulated with the SPH code \textsc{swift} (version 0.9.0) using the branch \texttt{planetary\_plus\_subtask\_speedup} (see Zenodo archive upon publication). Simulations were run using the cubic B-spline kernel (see kernel function $b_4$ in \cite{dehnen_aly_2012}, or Appendix~\ref{appendix:sim_data_table}). We used a \texttt{resolution\_eta} parameter (notated as $\eta$ in Section \ref{sect:sph_intro}) of 1.2348.

All simulations are of the same terrestrial giant impact scenario (shown in Figure~\ref{fig:giant_impact_example}), in which two identical proto-Earth-like planets collide, but with different code parameters. We varied the particle resolution used to represent each colliding object such that $N \in \{10^{5}$, $10^{6}$, $ 10^{6.5}$, $ 10^{7}\}$, and the maximum smoothing length such that $h_{\max} \in \{10^{-\frac{1}{2}} \rm R_{\oplus}$, $1 \rm R_{\oplus}$, $2 \rm R_{\oplus}$, $10^{\frac{1}{2}} \rm R_{\oplus} \}$, in order to investigate the effect of these simulation parameters on the results of our post-processing pipeline. The combinations of $N$ and $h_{\max}$ values used in this study, with the resulting minimum densities, are shown in Figure \ref{fig:impact_suite}.

In all simulations, both the target and impactor had a radius of $\sim 1 \rm R_{\oplus}$ and a mass of $\sim 0.9 \rm M_{\oplus}$, although there were some small variations with resolution due to the initialisation method. For exact radius and mass values, see Table~\ref{tab:all_sim_data}. Each object was modelled as having two layers - a metal core (containing $\sim 30 \%$ of the mass) and a rocky mantle (containing $\sim 70 \%$ of the mass). To model the core and mantle, we used a modified ANEOS equation of state \citep{Stewart_2020_eos_changes} with parameters for pure iron \citep{pure_iron_aneos_sarah_t_stewart} and forsterite \citep{forsterite_aneos_stewart+2019,Stewart_2020_eos_changes}, respectively. Neither the target nor the impactor were rotating prior to the impact. 

The radial thermal profile of our (identical) target and impactor were generated with WoMa \citep{Ruiz_Bonilla_2020}, and translated into SPH particles
via SEAGen \citep{kegerreis+19_seagen}. The thermal profile used in this study was set by fixing the surface at a pressure of 10~GPa and temperature of 2000~K. The thermal profiles of the core and mantle are both isentropic, with fixed entropies of 1900~$\rm J \ kg^{-1} \ K^{-1} $ and 2800~$\rm J \ kg^{-1} \ K^{-1} $ respectively. 

Before simulating collisions of these initialised SPH representations of planets, each initialised planet was settled via a 10\,h evolution in \textsc{swift} to ensure the planet had reached an equilibrium structure, and that any equilibrium velocities were damped. For the first 5\,h of settling, the entropy of the core and mantle were imposed to ensure that the resulting planetary thermal profile was as originally described. This pre-impact stabilisation used a simulation box size of $100 \ \rm R_{\oplus}$.

In each impact simulation, the target and impactor were placed at initial positions such that the point of impact occurred at $\sim$1\,h into the simulation, to allow pre-impact tidal deformation to take place. Our identical impact scenarios used an impact parameter of $b = \sin \theta = 0.7$, corresponding to an impact angle of $\theta \approx 45^{\circ}$. Each collision was initialised such that the velocity of the impactor relative to the target at the point of collision, $v_c$, was the mutual escape velocity of the target and impactor, $v_{\rm esc} \approx 10.6 {\ \rm km \ s^{-1} }$ (see Table~\ref{tab:all_sim_data} for all exact values of $v_c$).

The simulation data reported in this work was evolved until a simulation time of 24\,h, or 23\,h after the point of impact. Impact simulations used a box size of $1000 \, \rm R_{\oplus}$, and no particle had left the box in any of our colliding systems after 24\,h of simulation time.

\subsection{Determining Optical Thickness}
\label{subsect:ot_theory}

In order to locate the optically thick region of an impact structure, viewed at a given orientation, we need to consider the probability, $\mathcal{P}$, of a photon being absorbed whilst traversing a section of that structure. Calculating absorptivity is complicated by the fact that the evolving material is a combination of multiple phases (solid dust, liquid droplets, vapour, and potentially even plasma) and different chemical compositions.

Our ability to solve this complete problem is limited by the fact that current SPH impact simulations are not able to include thermal, chemical or material exchange between particles. Each particle in a simulation is assigned a material type which is retained from start to finish (e.g., in simulations conducted as part of this study, particles are either tagged as pure iron or pure forsterite) and treated as pseudo single component materials (e.g. with congruent melting and vaporisation). However, within these limitations, the following approach allows us to take into account the effect of multiphase material and simple mixtures of materials in impact structures. In this preliminary study, we are not attempting to calculate full radiative transfer. Instead, we focus on primary absorption/scattering of incoming/outgoing radiation.

To calculate $\mathcal{P}$ for a system consisting of a mixture of condensates (liquid droplets and/or dust in vapour), we follow the same approach taken by \cite{lock2018_moon}, who express $\mathcal{P}$ as:

\begin{equation}
    \label{eqn:photon_prob}
    \mathcal{P}(r) = \alpha(r){\rm d}r = \frac{{\rm d}r}{L_{\rm MFP}(r)} + \alpha_{\rm vap}(r){\rm d}r \; .
\end{equation}

\noindent Here, $\alpha(r)$ is the combined (condensate and vapour) absorption coefficient at the location described by $r$, $L_{\rm MFP}(r)$ is the mean free path of a photon travelling through a condensate suspension, and $\alpha_{\rm vap}(r)$ is the average absorption coefficient for vapour. The two terms on the right hand side of Equation \ref{eqn:photon_prob} are individually responsible for absorption due to condensates and vapour, respectively. Note that this absorptivity metric effectively assumes that vapour is the major phase by volume at the locations it evaluates, which is justified by the fact that we see very low condensate volume fractions in the outer regions of our structure (see Figure \ref{fig:cond_frac_vol_example}). If we integrate $\mathcal{P}(r)$ over the path that a photon takes through the impact structure, we can consider the structure to be optically thick where
\begin{equation}
    \label{eqn:integrate_ot}
    \tau = \int \mathcal{P}(r') {\rm d}r' \geq 1
\end{equation}
along a given line of sight.

$L_{\rm MFP}(r)$ determines absorption due to condensates. For a uniform size of spherical condensates,
\begin{equation}
    \label{eqn:mean_free_path}
    {L_{\rm MFP}(r)} = \frac{2D_0}{3}\frac{V_{\rm tot}}{V_{\rm cond}}
\end{equation}
where $D_0$ is the diameter of the spherical condensates, and $V_{\rm cond}/V_{\rm tot}$ is the fraction of the local volume taken up by condensate \citep[see e.g.][]{lock2018_moon}. In effect, this description assumes that scattering off the surfaces and interiors of small condensates makes them an effective barrier to photons travelling in a specified direction. The range and distribution of condensate size is uncertain, as these droplets could variably be sourced from the break-up of melt by shear or vapour bubble nucleation, and/or the condensation of vapour. These condensate formation processes are not captured by SPH simulations. However, Equation \ref{eqn:mean_free_path} demonstrates that opacity of the mixed-phase system depends strongly on the value of $D_0$, with photons more likely to be blocked when passing through a given mass of fixed-density condensate when that condensate is made up of more droplets with smaller diameters. This is due to a simple inverse proportionality relation between the cross-sectional area of all droplets and $D_0$. As a default, we assume a constant condensate droplet size throughout this study, with $D_0 = 1 \rm \, cm$. Larger droplets are not likely to be coupled to vapour at the edge of post-impact bodies \citep{lock2018_moon}, and so this realistically represents a lower bound on opacity.

To determine $\alpha_{\rm vap}$, the absorptivity due to vapour, we use a semiconducting Drude model for silica fluid absorption as constrained by \cite{kraus2012_vap}. This bolometric absorptivity is defined by the following equation:
\begin{equation}
    \label{eqn:vapour_absorption}
    \alpha_{\rm vap} = B_0 \bigg(\frac{\rho_{\rm vap}}{\rho_n}\bigg)^{\frac{1}{3}} \bigg(\frac{T_{\rm vap}}{T_n}\bigg) e^{-\left(B_1 \frac{T_n}{T_{\rm vap}}\right)} e^{-\left(B_2 \frac{\rho_{\rm vap} \ T_n}{\rho_n \ T_{\rm vap} }\right)} \; ,
\end{equation}
where $T_n$ and $\rho_n$ are arbitrary constants used to scale vapour temperature, $T_{\rm vap}$, and vapour density, $\rho_{\rm vap}$. $B_0$, $B_1$ and $B_2$ are fitting parameters which accompany the choice of $T_n$ and $\rho_n$. In \cite{kraus2012_vap} and in this work $T_n = 4150\ \rm{K}$, $\rho_n = 1900\ \rm{kg}\ \rm{m^{-3}}$, $B_0 = 10^{17} \ \rm{m^{-1}}$,  $B_1 = 37$ and $B_2 = -11.6$. This determination of absorptivity averages over all wavelengths. This theoretical model is constrained by absorption coefficients for $\rm{SiO_2}$ fluid generated by the code VASP \cite[Vienna Ab initio Simulation Package,][]{vaspcode_blochl1994}. \cite{kraus2012_vap} used VASP to calculate absorbances for 5 density-temperature pairs (see Figure 5 of \cite{kraus2012_vap} for list), using 4150K and 5400 K as temperature values, and densities in the range 9.2 - 1900 $\rm{kg \ m^{-3}}$. Although the temperature and density pairs used to constrain the metric in Equation \ref{eqn:vapour_absorption} do not span the full range of temperatures and densities likely needed to locate photospheres, the theoretical underpinnings of the model (rather than, say, contriving a fitted functional form) make it more likely to be reliable outside of this temperature-density range. Further information on the derivation of this metric can be found in the Supporting Information of \cite{kraus2012_vap}.

Note that instead of using an absorptivity metric for forsterite vapour, or vapour of a forsterite-iron mix, we are using an absorptivity metric for silica vapour. This choice was made as the authors are not aware of a high-quality absorptivity equation for forsterite, other silicate end-members, or silicate-metal mixes, that can span the wide thermodynamic range required for our calculations. Furthermore, the speciation of silicates at the high temperatures of post-impact bodies can be very uncertain \citep{lock2018_moon,Fegley+23}. Here we rely on the Kraus silica vapour absorptivity equation as a proxy for a generic silicate, effectively assuming that iron is a minor and miscible \citep{lock2018_moon} component in most parts of the optically thin regions of the colliding system, as we confirm post-hoc. In Appendix~\ref{appendix:metric_comparison}, we discuss two other absorptivity metrics that have been used to study planetary bodies where the opacity of silicate vapour plays a critical role, examine their applications and compare them to \cite{kraus2012_vap}.

\begin{figure}
    \includegraphics[width=\linewidth]{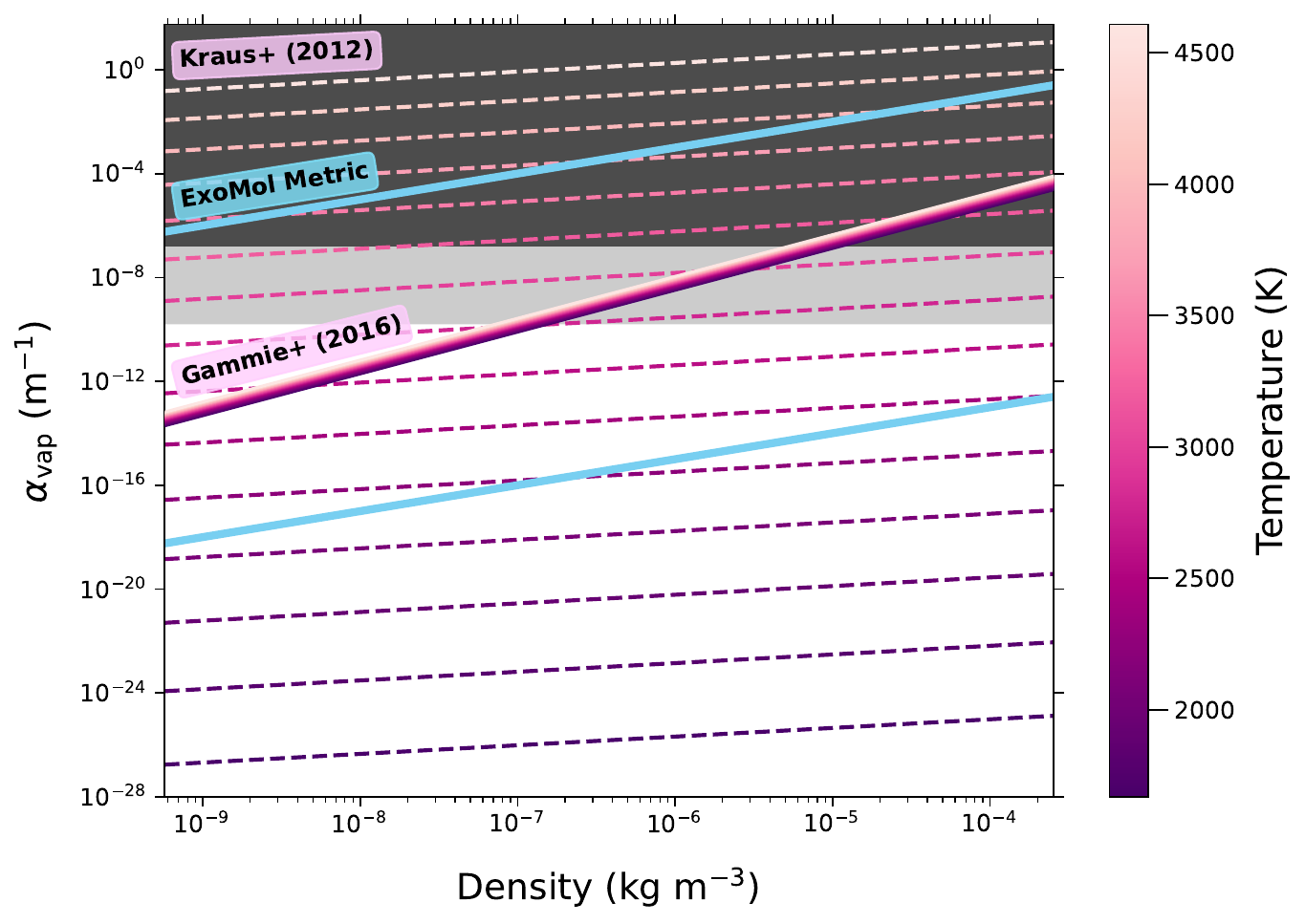}
    \caption{Visualising absorptivity metrics that could potentially be used to assess silicate vapour opacity. For each metric, absorptivity, $\alpha_{\rm{vap}}$, is plotted as a function of vapour density and colour-coded by vapour temperature. We compare Equation \ref{eqn:vapour_absorption} \protect\citep{kraus2012_vap}, shown as dashed lines, Equation \ref{eqn:park_opacity_eqn} \protect\citep{Gammie_2016}, shown as solid lines, and Equation \ref{eqn:exomol_vapour_absorption} (simple ExoMol metric), whose upper and lower bounds for absorption are shown as solid blue lines. Background colours indicate different optical thickness regimes: white indicates vapour absorptivity values for which a photon could travel the length of our SPH simulation box ($1000 \ \rm R_{\oplus}$) and not be absorbed; light grey indicates that a photon could travel a distance of $1 \rm{R_{\oplus}}$ without being absorbed but would certainly be absorbed if it travelled across the entire box; and dark grey indicates that a photon would certainly be absorbed if it travelled a distance of $1 \rm{R_{\oplus}}$.}
    \label{fig:absorptivity_metric_comparison}
\end{figure}

\subsection{Interrogation and Post-Processing of Simulated Data}
\label{subsect:data_processing}
\begin{figure}
    \includegraphics[width=\linewidth]{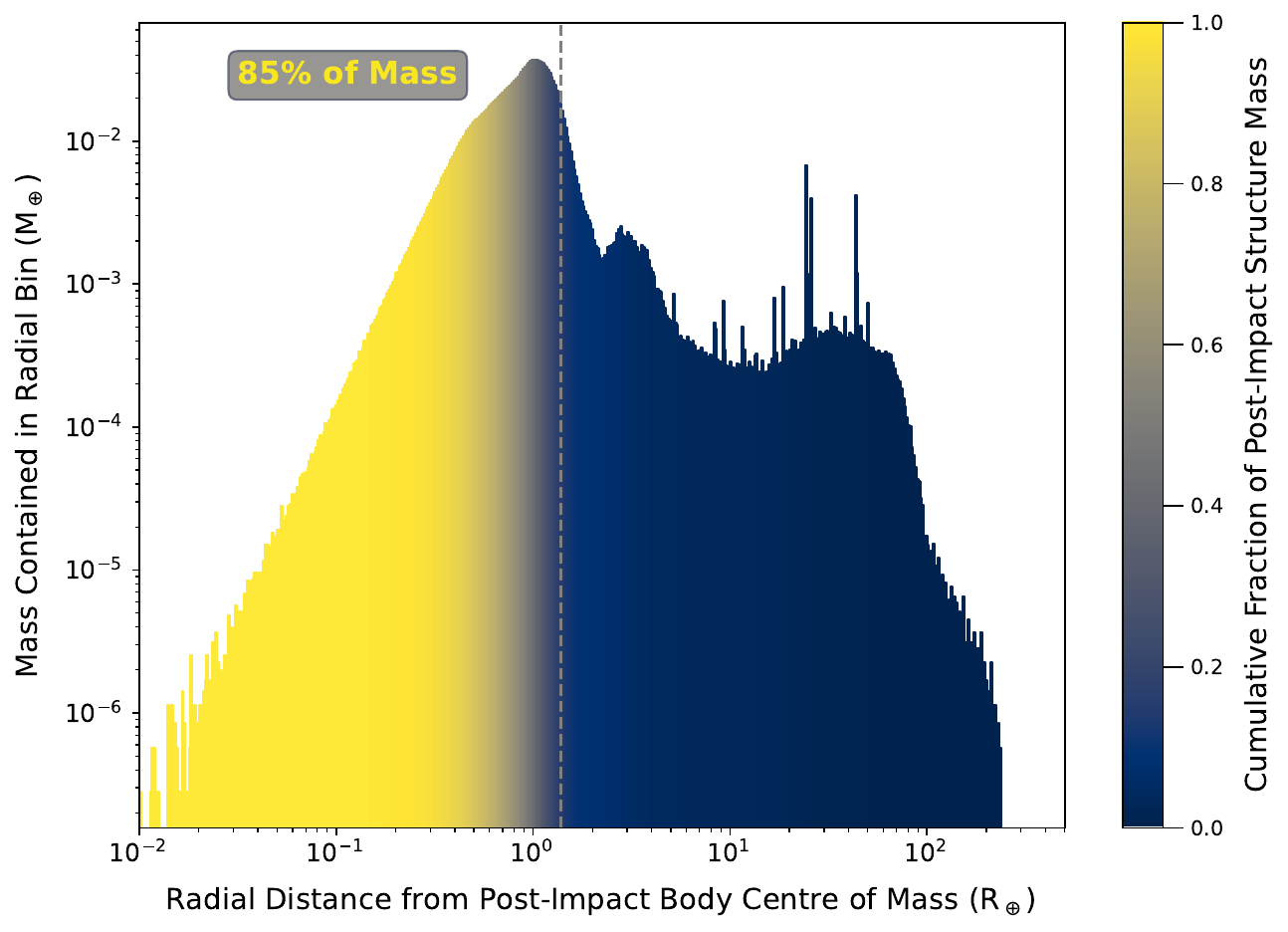}
    \caption{Mass distribution of an impact structure resulting from a simulated oblique collision between two proto-Earth-like planets, each resolved to approximately $10^{6.5}$ particles. The cumulative fraction of the impact structure's mass, calculated radially inwards, is used to colour-code this plot. The grey dotted line represents the radial separator between the innermost $85\%$ of mass and the outer regions of the body. The location of this separator tells us that $85\%$ of the impact structure's mass is contained within a radius of just a few Earth radii, even though the impact structure is a very extended object, highlighting the importance of accurately modelling its outer regions.}
    \label{fig:mass_dist_cumulative}
\end{figure}

We now consider how to apply the concepts discussed in Sections \ref{sect:sph_intro} and \ref{subsect:ot_theory} to the problem of analysing the observability of impact structures. The outer layers of impact structures are expected to constitute a very small portion of total mass, with mass being highly concentrated in the central bound remnant (e.g., Figure \ref{fig:mass_dist_cumulative}, where $\sim85\%$ of the mass is concentrated within a central region spanning a few Earth radii). For giant impacts similar to the example impact between two proto-Earth-like planets (Figure~\ref{fig:impact_suite}), the highest-resolution, largest $h_{\rm max}$ simulations that are typically computationally feasible, the smallest particle density that can be resolved is above ${10^{-5} \ \rm kg \ m^{-3}}$. As illustrated in Figure~\ref{fig:absorptivity_metric_comparison}, for temperatures above $\sim$ 3075 K, a purely vapour region of constant temperature and a constant density ${10^{-5} \ \rm kg \ m^{-3}}$ will become optically thick along a line of sight spanning just $1 \rm R_{\oplus}$. This calculation shows that the photic surface of impact structures is at very low densities (we will show that our example impact structures become optically thick at densities on the order of 10$^{-10}$ to $10^{-5}$~kg~m$^{-3}$ at 23~h after the impact). Therefore, we cannot simply use the raw output of SPH simulations to ascertain the size and thermal properties of the photic surfaces of impact structures, .

To accurately map impact photospheres, it is necessary to resolve the density structure that is normally hidden below the minimum density limit of SPH simulations, and emulate the corresponding (unresolved) thermodynamic properties such as temperature and phase. This requires a method for robustly determining the density and thermal state of the multiphase, multicomponent (iron and forsterite) system at any point in the structure, potentially a long way from the nearest SPH particle. We present such a method (a post-processing pipeline) in this section.

In Section~\ref{subsect:data_processing:particles}, we discuss the reassignment of smoothing lengths post-hoc after relaxing the runtime $h_{\rm max}$. In Sections~\ref{subsect:data_processing:arbitrary}~\&~\ref{subsect:data_processing:checks}, we describe and then validate a ``pseudo-particle'' routine for determining the density and thermodynamic properties at any point in the structure. We will return to discuss the reliability of our pipeline and the dependence on code parameters in Section~\ref{sect:analysis}.

\subsubsection{Reassignment of (Uncapped) Smoothing Lengths}
\label{subsect:data_processing:particles}
\begin{figure*}
    \includegraphics[width=\textwidth]{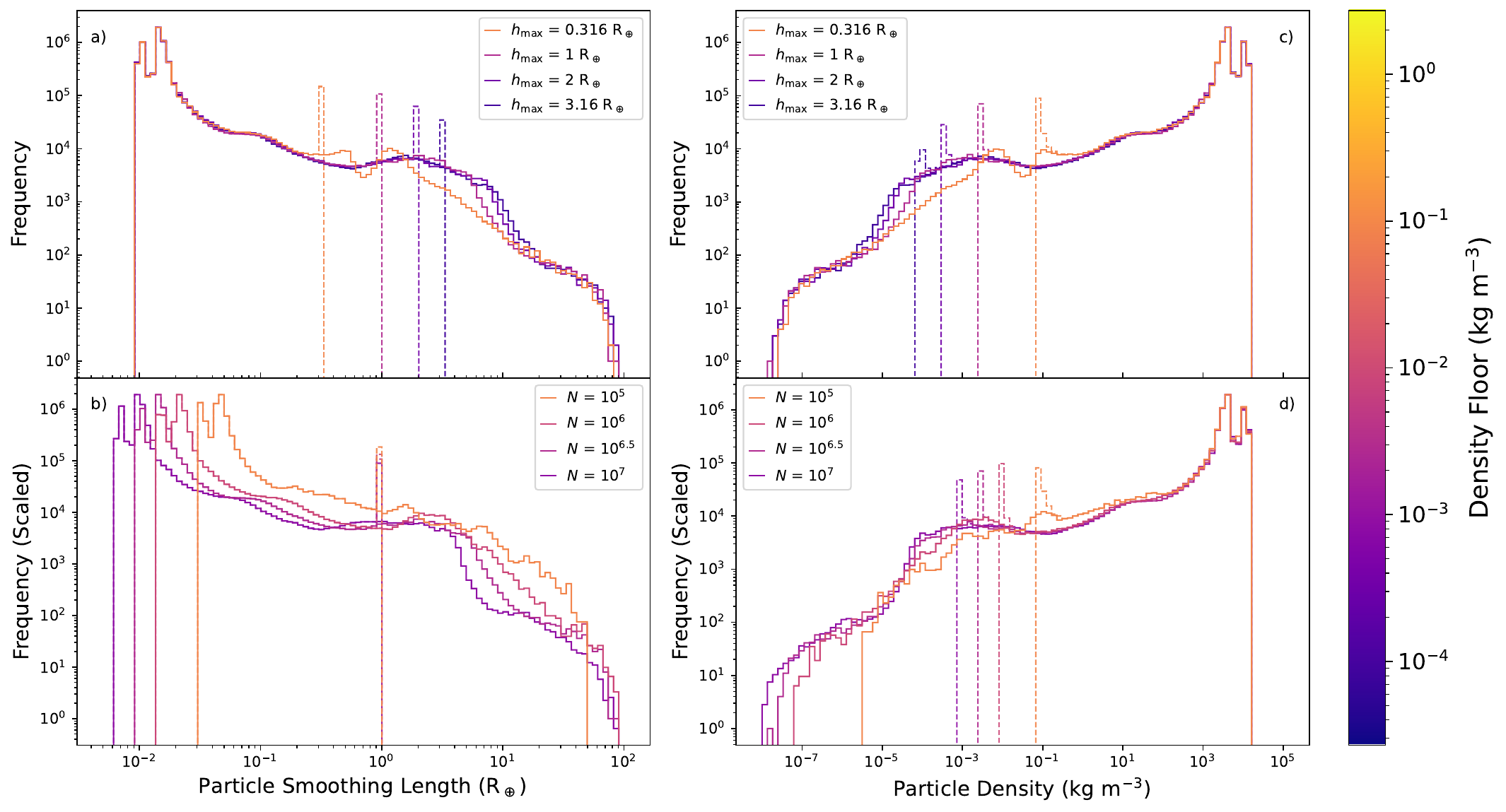}
    \caption{Histograms showing original interpolated particle densities when smoothing lengths are truncated at $h_{\rm max}$ (dotted lines), vs. reprocessed interpolated particle densities (solid lines), for a range of different values of $N$ and $h_{\rm max}$, colour-coded by simulation density floor. The original distributions are taken 24\,h (23h after the impact event) into simulations of our terrestrial giant impact example. The left-hand panels compare original and post-processed smoothing length distributions for a) simulations where $N = 10^{6.5}$, for 4 different values of $h_{\rm max}$, and b) simulations where $h_{\max} = 1 R_\oplus$, for 4 different resolutions (where counts have been rescaled to be comparable to the $N = 10^{6.5}$ case). Likewise, the right-hand panels compare original and post-processed densities (the latter being re-interpolated from new smoothing length values) for c) simulations where $N = 10^{6.5}$, for 4 different values of $h_{\rm max}$, and d) simulations where $h_{\max} = 1 R_\oplus$, for 4 different resolutions (again with rescaled counts). Note the large spikes in the original distributions, indicating either an imposed value of $h_{\max}$ (a and b) or the resulting density floor (c and d). The original and reprocessed smoothing length distributions are identical when $h < h_{\rm max}$. However, the original and post-processed distributions are not identical above the density floor, since it is possible to be at $h_{\rm max}$ and not be isolated. Note that in the case that $h_{\rm max} < 1R_\oplus$, there is a build-up of particles with smoothing lengths just larger than $h_{\rm max}$, indicating that values $h_{\rm max}$ that are too low cause particles to clump together, something that location-based post-processing is not able to correct for.}
    \label{fig:reassign_rho}
\end{figure*}

Throughout a simulation, \textsc{swift} assigns smoothing lengths to particles by iteratively solving Equations \ref{eqn:sph_iterative_h_solve_a} \& \ref{eqn:sph_iterative_h_solve_b}, but for particles for which the solution would be above the user-specified value of $h_{\rm max}$, the solution is truncated (see Section~\ref{sect:sph_intro}). The effect of this truncation can be seen in histograms of the smoothing lengths and densities of particles in our simulations (dashed lines, Figure~\ref{fig:reassign_rho}).

In order to emulate realistic density values in regions of the structure that have hit $h_{\rm max}$, we replicate the iterative smoothing length assignment performed by \textsc{swift} during runtime \cite[as described in][]{Price_2012} on simulation output, but do not truncate at $h_{\rm max}$. To achieve this, we first generate a tree for nearest-neighbour lookup from the SPH particle position data via the \texttt{KDTree} function of the \texttt{scipy.spatial} package (\texttt{scipy} version 1.11.4). This tree can then be easily queried to ascertain either the $n$ nearest neighbours to a given location, or how many SPH particles are within a given radius of that location. Note that if the position queried is itself the central location of an SPH particle, the particle will be considered one its own neighbours (consistent with SPH interpolation).

To reassign the smoothing length of an SPH particle, we iteratively solve Equations~\ref{eqn:sph_iterative_h_solve_a} \& \ref{eqn:sph_iterative_h_solve_b}, with the aid of the particle position tree. To provide an initial guess, we first find the $[N_{\rm neigh}]~+~1$ nearest neighbours to our particle's location (this list will include itself). In our simulation set-up, $N_{\rm neigh} \approx 48$ (see Equation \ref{eqn:neighbour_number}). We set an initial guess for $h$ that corresponds to having a kernel support radius $H$ extending out to the midpoint of the distances to the $48^{\rm th}$ and $49^{\rm th}$ nearest neighbours. We arrive at a solution iteratively using the \texttt{fsolve} and \texttt{root} functions of the \texttt{scipy.optimize} package. Once we have solved for the unrestricted smoothing length, we can interpolate an updated value for density, $\rho$, using Equation \ref{eqn:sph_rho}. The recalculation of smoothing lengths extends the range of densities being considered by several orders of magnitude, as shown by the solid lines in Figure~\ref{fig:reassign_rho}.

To recover more realistic thermodynamic states, we conceptually assume that each $h_{\rm max}$ particle expanded isentropically to the volume dictated by its reassigned smoothing length. SPH particles behave adiabatically regardless of changes to their smoothing lengths, except when shocked or sheared, and SPH impact codes do not consider radiative transfer \citep{Price_2012}. Therefore, entropy provides a reliable tracer of a particles thermodynamic state, save for physical processes not captured by the code. Other thermodynamic properties (temperature, pressure, etc.) can then be recalculated using $\rho$ and $s$ and the EoS for the material that makes up that particle (see Section~\ref{subsect:giant_impact_how_to}). Changes in specific internal energy and temperature effectively reflect the reversible thermodynamic ($P\mathrm{d}V$) work that would have been done in expanding the particle isentropically to lower densities.

In following this method for recalculating the thermodynamic parameters, we effectively assume that the $h_{\rm max}$ particles are expanding but doing no net work on each other. In reality, pressure forces between the parcels of material represented by the SPH particles would have led to additional accelerations of the outer portions of the structure not captured in our simulation or post-processing. The extent of any impact bodies we calculate will therefore likely be underestimated, to a degree dependent on the choice of simulation resolution and maximum smoothing length (see Section~\ref{sect:analysis}).

\subsubsection{Interpolating Properties at Arbitrary Locations}
\label{subsect:data_processing:arbitrary}
\begin{figure*}
    \centering
    \includegraphics[width=\textwidth]{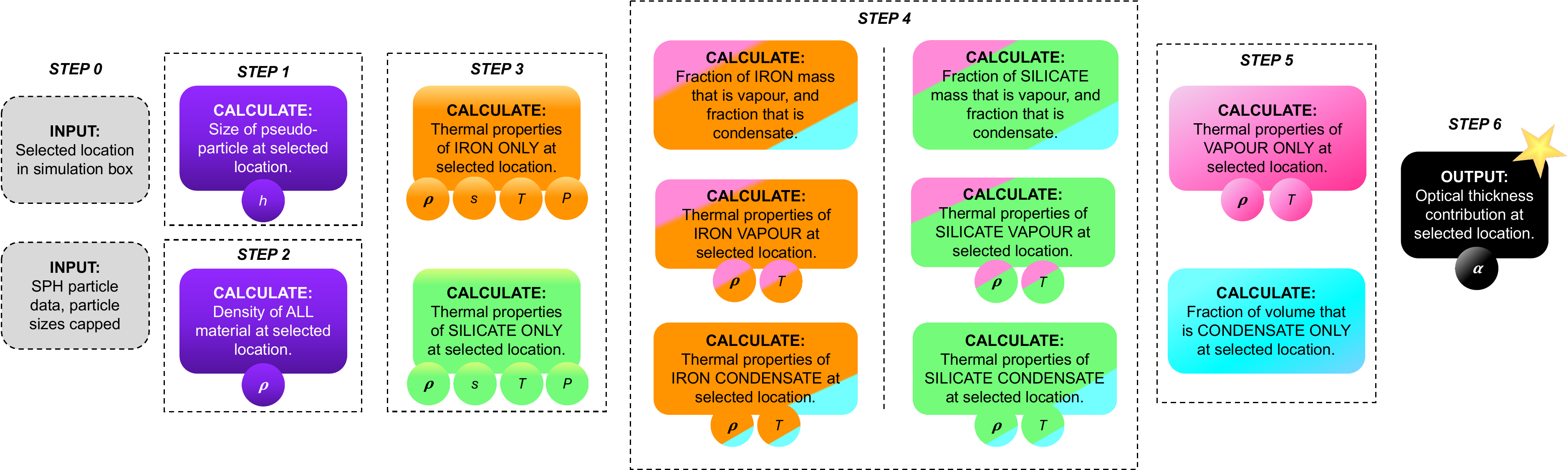}
    \caption{Schematic of absorptivity evaluation algorithm to be used in the determination of optical thickness, as described in Section \ref{subsect:data_processing}. The aim of the algorithm is, given SPH simulation output from a giant impact and an arbitrary location in the simulation box, to determine $\frac{V_{\rm cond}}{V_{\rm tot}}$, $\rho_{\rm vap}$ and $T_{\rm vap}$, in order to calculate absorptivity $\alpha$ (see Equations \ref{eqn:photon_prob}, \ref{eqn:mean_free_path}, \ref{eqn:vapour_absorption}). Grey represents algorithm input, black represents algorithm output. Steps involving isolating iron or silicate components are orange or green respectively. Steps involving isolating vapour or condensate components are pink or blue respectively. Other stages of this algorithm (that deal with all materials and all phases at once) are purple.}
    \label{fig:algo_schematic}
\end{figure*}

To map the photosphere of an impact structure, we need to be able to calculate thermodynamic properties at any arbitrary location in the simulation box rather than only having information at the centres of SPH particles, in order to integrate absorptivity along lines of sight. To do this, we apply a post-processing pipeline, a schematic of which is shown in Figure~\ref{fig:algo_schematic}.

To begin with, we follow a process nearly identical to that used when considering the reassignment of smoothing lengths for individual SPH particles (Section~\ref{subsect:data_processing:particles}), but instead of SPH particles, we consider ``pseudo-particles''. At any point in the simulation box, irrespectively of whether it is the central location of an SPH particle, we can consider there to be a psuedo-particle, and thus assign an uncapped proxy smoothing length to that location (Step 1, Figure~\ref{fig:algo_schematic}). We solve Equations \ref{eqn:sph_iterative_h_solve_a} \& \ref{eqn:sph_iterative_h_solve_b} as for individual particles. Once we have solved for a proxy value of $h$ for our pseudo-particle, we can interpolate a value for density, $\rho$, using Equation \ref{eqn:sph_rho} (Step 2, Figure~\ref{fig:algo_schematic}). 

Determining the thermodynamic properties other than density for an arbitrary location (Steps 3--6) is more complicated, due to the fact that the impact structure is multicomponent and multiphase. SPH particles represent parcels of mass made up of a single material; no material mixing, nor chemical and thermal exchange is taken into account during the runtime of the simulation. When recalculating the thermal behaviour of SPH particles, we use the EoS for its material (in our case, forsterite or iron). However, thermodynamic properties at arbitrary locations in the structure will be influenced by multiple particles which may be of different materials. We therefore devised an interpolation method that accounts for the contributions from SPH particles of different materials separately. This allows us to determine the contribution of individual materials to the density and thermal properties at any point in the structure (Step 3). We can then apply the EoS of each material to determine the vapour and condensate mass fraction of each material, and their properties (Step 4). Finally, we determine the properties of the mixed-material vapour and mixed-material condensate (Step 5) to be used in calculating the opacity at our given location (Step 6). 

Although our pseudo-particle method can be used to interpolate material properties at any arbitrary location in our simulation box, that is not to say that it should be. It is important to employ this method only at locations for which there is enough information from the SPH particles to approximate the density and thermodynamic properties of the system. We therefore define an approximate ``information region'' to be the smallest sphere, centred on the centre of mass, that would contains the recalculated kernels of all SPH particles. In this study, for a condensate droplet diameter of $D_0 = 1 \rm \, cm$, we locate photospheres well within our defined information regions (see Section \ref{subsect:discussion:limitations_SPH} for further discussion). Results involving significant interpolation outside of the information region will not be representative of the real system, and may unintentionally introduce extra mass into the system, or violate other conservation laws.

Steps 3--5 are implemented as follows: After calculating the smoothing length and density of a pseudo-particle at a location $\mathbf{x}$ (Steps 1~\&~2), we interpolate the specific entropies of the SPH particles contained within that pseudo-particle volume, separately by particle material type. The interpolation is weighted by a kernel with the smoothing length calculated in Step 1. In line with Equation~\ref{eqn:sph_interpolant}, we define $A(\mathbf{x})$ as iron entropy averaged over pseudo-particle volume, such that
\begin{equation}
    A_j = \begin{dcases}
        \frac{m_j s_j}{V_j} &\text{if $j$ denotes an iron particle} \\
        0 &\text{otherwise}
    \end{dcases}
\end{equation}
where $s_j$ denotes the specific entropy of the SPH particle $j$. Similarly, we define $B(\mathbf{x})$ as iron mass averaged over pseudo-particle volume, such that
\begin{equation}
    B_j = \begin{dcases}
        \frac{m_j}{V_j} &\text{if $j$ denotes an iron particle} \\
        0 &\text{otherwise.}
    \end{dcases}
\end{equation}
Note that $B$ is not the same as the individual density, $\rho_{\rm fe}$, since we are averaging iron mass over the pseudo-particle volume determined by proxy smoothing length $h$, rather than averaging iron mass over the volume occupied by the iron mass. Using these definitions of $A$ and $B$, we can define the specific entropy of iron at location $\mathbf{x}$ as:
\begin{equation}
    \label{eqn:interp_s_fe}
    s_{\rm fe} = \frac{S_{\rm fe}}{V_{\rm tot}} \cdot\frac{V_{\rm tot}}{m_{\rm fe}} = \frac{A(\mathbf{x})}{B(\mathbf{x})} = \frac{\sum_{k} \ m_k s_k \ W(|\mathbf{x - \mathbf{x}}_k|, h)}{\sum_{k} \ m_k \ W(|\mathbf{x - \mathbf{x}}_k|, h)}
\end{equation}
where the index $k$ denotes summing over iron particles only, and $s_k$ is the specific entropy of the SPH particle $k$.
Using the same logic for forsterite, we obtain:
\begin{equation}
    \label{eqn:interp_s_sil}
    s_{\rm fors} = \frac{\sum_{l} \ m_l s_l \ W(|\mathbf{x - \mathbf{x}}_l|, h)}{\sum_{l} \ m_l \ W(|\mathbf{x - \mathbf{x}}_l|, h)}
\end{equation}
where the index $l$ denotes summing over silicate SPH particle contributions only. Extending this logic, we can also approximate the mass fractions of iron and silicate at location $\mathbf{x}$ as:
\begin{equation}
        \begin{aligned}
        f_{\rm fe} & = \frac{m_{\rm fe}}{m_{\rm tot}} \approx  \frac{\sum_{k} \ m_k\ W(|\mathbf{x - \mathbf{x}}_k|, h)}{\sum_{j} \ m_j \ W(|\mathbf{x - \mathbf{x}}_j|, h)} \\
        f_{\rm fors} & = \frac{m_{\rm fors}}{m_{\rm tot}} \approx \frac{\sum_{l} \ m_l\ W(|\mathbf{x - \mathbf{x}}_l|, h)}{\sum_{j} \ m_j \ W(|\mathbf{x - \mathbf{x}}_j|, h)}
    \end{aligned}
\end{equation}
Therefore, Equations \ref{eqn:interp_s_fe}~\&~\ref{eqn:interp_s_sil} are consistent with the SPH interpolation of the specific entropy of the pseudo-particle, since:
\begin{equation}
    \label{eqn:total_entropy_consistent}
    s_{\rm tot} = \frac{\sum_{j} \ m_j s_j\ W(|\mathbf{x - \mathbf{x}}_j|, h)}{\sum_{j} \ m_j \ W(|\mathbf{x - \mathbf{x}}_j|, h)} = f_{\rm fe} s_{\rm fe} + f_{\rm fors} s_{\rm fors}.
\end{equation}
neglecting any entropy of mixing. In the case that all SPH particles within the pseudo-particle volume are iron, we use $s_{\rm fe}$ = $s_{\rm tot}$, and likewise that $s_{\rm fors}$ = $s_{\rm tot}$ if all particles are forsterite. Furthermore, it is important to note that if $s_{\rm tot}$ is evaluated at the central location of an SPH particle with specific entropy $s_i$, the interpolation will not retrieve $s_{\rm tot} = s_{i}$, because a) the $s$ we calculate is an average over the pseudo-particle volume so not directly comparable to a single particle, and b) we are not defining $s_{\rm tot}$ as a single function akin to Equation~\ref{eqn:sph_interpolant}, but rather the ratio of two such functions.

In effect, the material at a given point in the structure is assumed to be a mechanical mixture of the two materials, with no chemical or thermal exchange, and no contribution to entropy from thermodynamic mixing. In reality, silicates and iron can be miscible in large volumes of the collisional structures \citep{lock2018_moon,caracas_stewart_2023,postema+2026} or undergo chemical exchange reactions. Accounting for the entropy of mixing (or enthalpy of reaction) in such regions would alter the combined entropy of the system, but without a complete, wide-ranging EoS for a (miscible) silicate-metal system it is not yet possible to take this effect into account. As in Section~\ref{subsect:data_processing:particles}, we take particle specific entropy values, $s_j$, as tracers of thermal state, and use these interpolated material-specific entropies to calculate other material-specific thermal properties.

As density is not a simple additive property, in calculating the individual densities of separate materials at a given point, it is necessary to make an additional assumption. Here, we require the pressure of the metal and silicate to be in equilibrium, as well as having density values that are consistent with the total density calculated for the pseudo-particle at that point. Now that we have calculated specific entropies for individual materials at $\mathbf{x}$ (Equations~\ref{eqn:interp_s_fe} and \ref{eqn:interp_s_sil}), we iteratively solve the following coupled equations to obtain $\rho_{\rm fe}$ and $\rho_{\rm fors}$:
\begin{gather}
    \label{eqn:pressure_balancing}
     P_{\rm fe}(\rho_{\rm  fe}, s_{\rm fe}) = P_{\rm fors}(\rho_{\rm  fors}, s_{\rm fors}) \\
     \label{eqn:density_balancing}
    \frac{\sum_k m_k W(|\mathbf{x - \mathbf{x}}_k|, h)}{\rho_{\rm fe}} + \frac{\sum_l m_l W(|\mathbf{x - \mathbf{x}}_l|, h)}{\rho_{\rm fors}} = 1 \; .
\end{gather}
where we have used $f_{\rm fe} \rho_{\rm tot} = \sum_k m_k W(|\mathbf{x - \mathbf{x}}_k|, h)$ and the analogous expression for $f_{\rm fors} \rho_{\rm tot}$.
Our iterative solving procedure uses $\rho_{\rm fe} = \rho_{\rm fors} = \rho_{\rm tot}$ as an intial guess for Equation~\ref{eqn:pressure_balancing}. We use the EoS for each material to calculate pressures, in tandem with the SPH particle \texttt{KDTree} and the \texttt{fsolve} function from \texttt{scipy.optimize} package. Note that if all particles within a pseudo-particle volume are of the same material, we instead use $\rho_{\rm fe} = \rho_{\rm tot}$ or $\rho_{\rm fors} = \rho_{\rm tot}$ (as appropriate).

Once we have obtained density, specific entropy and pressure values for each material contribution at location $\mathbf{x}$, we can then use the EoS of each material to find the temperatures that correspond to each material, $T_{\rm fe}$ and $T_{\rm fors}$, thus completing Step 3. Note that there is no requirement for $T_{\rm fe}$ and $T_{\rm fors}$ to be equal.

We can then use the material-specific temperature-density pairs in conjunction with the condensate-vapour phase boundary of the relevant EoS to evaluate the phase of the iron and forsterite contributions. For our impact structures, we find that either material is pure vapour, pure condensate, or a mixture of these two phases. In the case that the material is mixed phase, we determine the material-wise mass fractions of vapour and condensate using the lever rule (Step 4):
\begin{gather}
    \label{eqn:iron_lever_rule}
    \frac{1}{\rho_{\rm fe}} = \frac{m_{\rm fe, cond}}{m_{\rm fe}} \frac{1}{\rho_{\rm fe, cond}(T_{\rm fe})} + \frac{m_{\rm fe, vap}}{m_{\rm fe}} \frac{1}{\rho_{\rm fe, vap}(T_{\rm fe})} \\
    \label{eqn:fors_lever_rule}
    \frac{1}{\rho_{\rm fors}} = \frac{m_{\rm fors, cond}}{m_{\rm fors}} \frac{1}{{\rho_{\rm fors, cond}}(T_{\rm fors})} + \frac{m_{\rm fors, vap}}{m_{\rm fors}} \frac{1}{\rho_{\rm fors, vap}(T_{\rm fors})}  
\end{gather}
where $\rho_{\rm fe, cond}$, $\rho_{\rm fe, vap}$, $\rho_{\rm fors, cond}$ and $\rho_{\rm fors, vap}$ are the densities of iron condensate, iron vapour, forsterite condensate and forsterite vapour, respectively. These density values are obtained by looking up the density of material on the condensate and vapour sides of the phase boundary at temperature $T_{\rm fe}$ or $T_{\rm fors}$. We solve Equation~\ref{eqn:iron_lever_rule} for $m_{\rm fe, cond} /m_{\rm fe}$ and $m_{\rm fe, vap} /m_{\rm fe}$, respectively the fraction of iron mass that is condensate and the fraction of iron mass that is vapour. Similarly, we solve Equation~\ref{eqn:fors_lever_rule} for $m_{\rm fors, cond} /m_{\rm fors}$ and $m_{\rm fors, vap} /m_{\rm fors}$, respectively the fraction of silicate mass by condensate and the fraction of silicate mass by vapour. Here, we make use of the fact that $m_{\rm fe, cond}\ +\ m_{\rm fe, vap} = m_{\rm fe}$ and $m_{\rm fors, cond}\ +\ m_{\rm fors, vap} = m_{\rm fors}$. Therefore, at any location and for a terrestrial impact structure, we can obtain thermodynamic information for four categories of material: silicate vapour, silicate condensate, iron vapour and iron condensate. 

Note that equations of state for planetary materials are not especially well constrained in the high-entropy, very low density regions of the thermodynamic parameter space relevant to our study. The ANEOS routine for calculating the condensate-vapour phase boundary \citep{thompson_aneos_1990} can occasionally fail to find physical solutions to the phase boundary at low pressures, which require some minor corrections to avoid negative pressures and ensure that vapour and condensate pressures are approximately equal at a given temperature. Further, to avoid artificial discontinuities in radial vapour density profiles that can arise from undersampling phase transition regions in the EoS, we used a very finely-gridded EoS and phase boundary in our post-processing. All EoS tables and corresponding phase curves used in our analysis are available in the supporting material accompanying this manuscript.

From the properties of the individual phases of each material we can calculate the total (overall) condensate volume fraction, vapour density and vapour temperature at a given location in the simulation box (Step 5):
\begin{gather}
    \label{eqn:cond_terms_vol}
    \frac{V_{\rm cond}}{V_{\rm tot}} = f_{\rm fe} \ \frac{{m_{\rm fe, cond}}}{m_{\rm fe}} \ \frac{\rho_{\rm tot}}{{\rho_{\rm fe, cond}}} + f_{\rm fors} \ \frac{{m_{\rm fors, cond}}}{m_{\rm fors}} \ \frac{\rho_{\rm tot}}{{\rho_{\rm fors, cond}}} \\ 
    \label{eqn:vap_terms_rho}
    \rho_{\rm vap} = \bigg(\frac{1}{{\rho_{\rm fe, vap}}} \frac{{m_{\rm fe, vap}}}{m_{\rm vap}} + \frac{1}{{\rho_{\rm fors, vap}}} \frac{{m_{\rm fors, vap}}}{m_{\rm vap}} \bigg)^{-1} \\
    \label{eqn:vap_terms_T}
    T_{\rm vap} = \frac{{m_{\rm fe, vap}}}{m_{\rm vap}} {T_{\rm fe, vap}} + \frac{{m_{\rm fors, vap}}}{m_{\rm vap}} {T_{\rm fors, vap}}.
\end{gather}
All of the terms in Equations \ref{eqn:cond_terms_vol}, \ref{eqn:vap_terms_rho} and \ref{eqn:vap_terms_T} are previously obtained through following Steps 1--4, except for the vapour mass fractions, which we calculate from known quantities via
\begin{gather}
    \frac{{m_{\rm fe, vap}}}{m_{\rm vap}} = \frac{{m_{\rm fe, vap}}}{m_{\rm fe}} \frac{m_{\rm fe}}{m_{\rm tot}} \bigg( \frac{m_{\rm vap}}{m_{\rm tot}} \bigg)^{-1} \\
    \frac{{m_{\rm fors, vap}}}{m_{\rm vap}} = \frac{{m_{\rm fors, vap}}}{m_{\rm fors}} \frac{m_{\rm fors}}{m_{\rm tot}} \bigg( \frac{m_{\rm vap}}{m_{\rm tot}} \bigg)^{-1}
\end{gather}
where
\begin{equation}
    \label{eqn:vap_mass_frac}
    \frac{m_{\rm vap}}{m_{\rm tot}} = \frac{{m_{\rm fe, vap}}}{m_{\rm fe}} \ \frac{m_{\rm fe}}{m_{\rm tot}} + \frac{{m_{\rm fors, vap}}}{m_{\rm fors}} \ \frac{m_{\rm fors}}{m_{\rm tot}}.
\end{equation}
Being able to calculate $V_{\rm cond}/V_{\rm tot}$, $\rho_{\rm vap}$ and $T_{\rm vap}$ at any arbitrary location in our simulation box, we have all the information we need to integrate Equation \ref{eqn:integrate_ot} and locate the surface $\tau = 1$ (Step 6, see Section~\ref{subsect:photosphere_emission_method}).

Note that we do not need to calculate accurate thermal profiles of all SPH particles in order to interpolate absorptivity at arbitrary locations -- our post-processing requires solely particle locations, masses and specific entropies. The smoothing lengths of all particles at $h_{\rm max}$ are recalculated only to define our information region, and updated internal energies are only used to verify that our method does not violate the conservation of internal energy (see Section~\ref{subsect:data_processing:checks}).

\subsubsection{Validation of Post-Processing Methods}
\label{subsect:data_processing:checks}
\begin{figure}
    \centering
    \includegraphics[width=\linewidth]{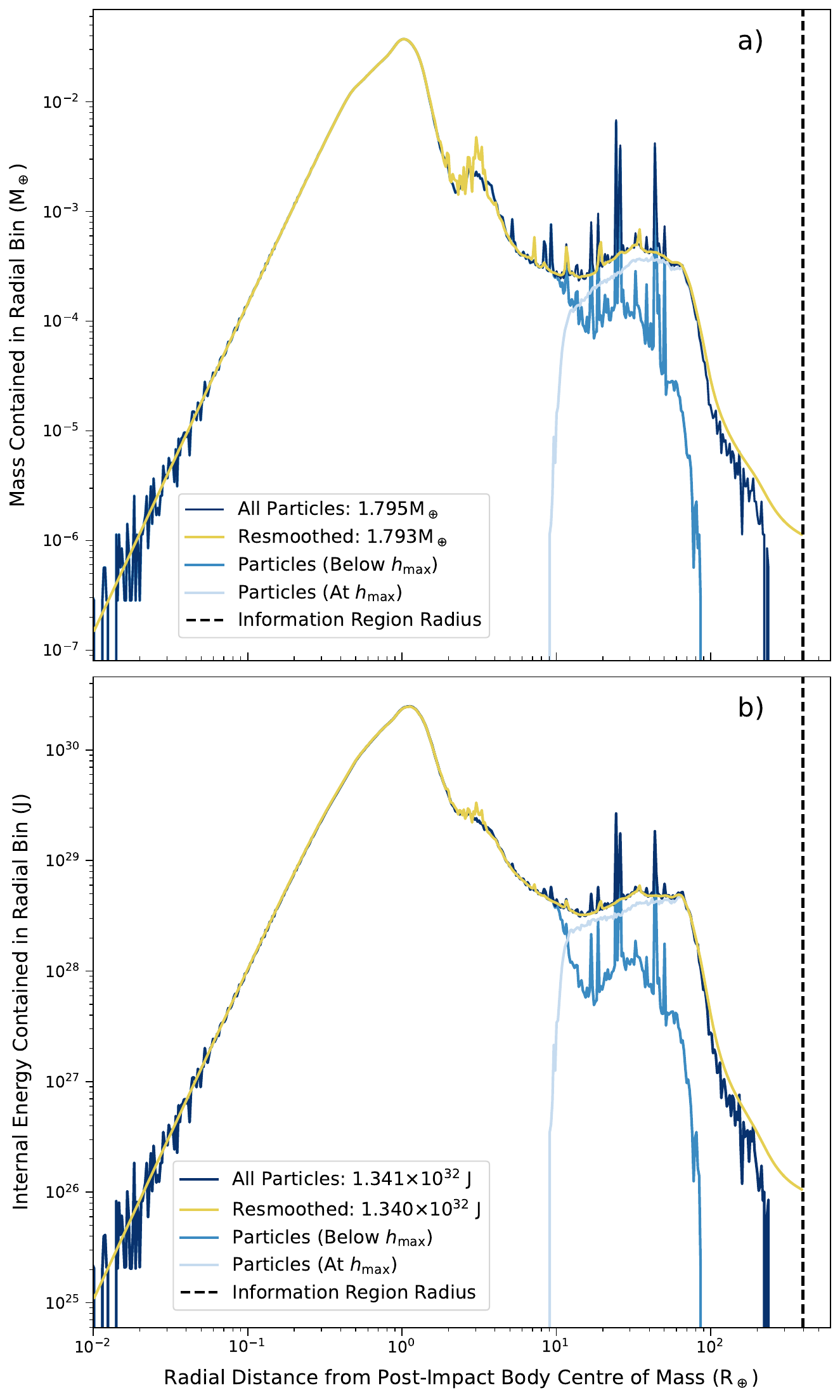}
    \caption{Verifying that our post-processing algorithm does not violate the mass and internal energy conservation for the terrestrial post-impact structure corresponding to $N \approx 10^{6.5}$ and $h_{\rm max} = 1 \rm R_{\oplus}$. We compare radially binned mass (panel a) and internal energy (panel b) distributions using two different methods: stored particle values of $m_i$ and reassigned values of $U_i$ (dark blue lines), versus radially binned smoothed masses and internal energies after post-processing interpolation (yellow lines). The legend reports the total mass or internal energy found via each method. For more information on how the radially binned smoothed mass and smoothed internal energy distributions are found, see Appendix~\ref{appendix:volume}. Binning by particle location is split into two sub-cases: particles that are below $h_{\rm max}$ (mid-blue) and particles that are at $h_{\rm max}$ (light blue). The black dotted line represents the radius at which the outer edge of the information region is located.}
    \label{fig:mass_dist_comparison}
\end{figure}

As intended, resmoothing masses and interpolating physical properties at arbitrary locations alters the mass distribution in the outer regions of impact structures from that which is described by the central locations of the SPH particles alone. Figure~\ref{fig:mass_dist_comparison}a shows an example comparison between a radial mass distribution based on the central positions of SPH particles (dark blue), and the interpolated mass distribution (yellow) plotted for the same radial bins. The interpolated mass distribution is calculated by approximating a volume integration of density over individual concentric spherical shells that make up the information region. The sampled density values in each shell are found using our pseudo-particle interpolation method. (See Appendix \ref{appendix:volume} for more details on the integration approximation.)

In the central region of the impact structure ($\lesssim 3 R_{\oplus}$), where most of the mass is contained, there is almost perfect agreement between the two methods beyond stochastic particle binning. This inner region is highly populated with SPH particles, no particles are isolated, and the reassigned smoothing lengths match the runtime values (Figure~\ref{fig:reassign_rho}). The consistency between the mass distributions in the inner region, therefore, demonstrates the robustness of our interpolation method, at least in an information-rich region. Beyond a few $\rm R_{\oplus}$, comparison between the two distributions becomes more complicated as the system becomes sparser. The particle mass distribution shows frequent, large spikes due to local clumps of particles (e.g., moonlets or spiral arms). The interpolated distribution is generally much smoother in the outermost regions of the system, where large pseudo-particle smoothing lengths smooth out the binned particle mass distribution. At a radial distance of about $50 R_\oplus$ from the impact structure centre of mass, the number of particles that have not reached the artificial smoothing cap ($h_{\rm max} = 1 R_\oplus$ in this case) begins to drop off very sharply, and beyond $90 R_\oplus$, all particles are isolated. It therefore may be expected that the two mass distributions start to diverge in the outermost regions of the structure, but this divergence is modest. 

As well as matching the general mass distribution, we verify that the mass of the total system within the information region is conserved by our post-processing interpolation routine. Integrating the post-processed mass distribution shown in Figure~\ref{fig:mass_dist_comparison}a over the information region, we recover 99.9\% of the total SPH particle mass. The example simulation being analysed involves $\sim 2 \times 10^{6.5}$ particles, and the mass difference between these approximations is equivalent to $\sim 6.1 \times 10^{3}$ particles. The ``missing'' mass is most likely due to our integration method (for mass per shell) not fully sampling the complete heterogeneity of the impact system (Appendix \ref{appendix:volume}).

We can also perform a similar check for the distribution and conservation of internal energy (also detailed in Appendix \ref{appendix:volume}). However, in this case, we post-process the individual SPH particles first, recalculating their internal energy after $h$ is reassigned, and then compare binning vs. interpolation (since work is effectively done expanding the SPH particles during post-processing). As with the mass distributions, we find very good agreement between particle-binned vs interpolated data, with similar stochastic fluctuations (Figure~\ref{fig:mass_dist_comparison}b). We recover 99.97\% of the total internal energy when integrating the interpolated distribution over the information region.

\subsection{Calculating Photospheric Temperature and Emission}
\label{subsect:photosphere_emission_method}
We are now able to integrate opacity along arbitrary lines of sight throughout our SPH-simulated structures, allowing us to determine the shape and thermal properties of the photospheres of impact structures. In particular, we wish to determine the effective temperature and luminosity of each photic surface.

It is very computationally expensive to precisely integrate absorptivity, due to the cost of interpolation in information-poor regions. We therefore approximate absorptivity $\alpha(r)$ along a given line of sight as a step function over intervals of equal length, $\Delta r$, giving $\tau$ at the $p$-th step along that line of sight as:

\begin{equation}
    \tau \approx \Delta r\sum_{q = 1}^{p} \alpha(r_q)
\end{equation}

To further reduce computational expense, we do not necessarily integrate over the whole information region. Instead, we only integrate over steps that contribute significantly to optical thickness. We begin with a 3D grid of sample points that encapsulates all particle locations, and add padding around that volume to ensure no significantly optically dense regions are excluded. In our cases, the unpadded grid of sample points is chosen to have resolution $101\times101\times101$, and lines of sight use $\Delta r$ in the range $\sim$~4.5--6~$\rm R_{\oplus}$. For this study, we only consider lines of sight that are perpendicular to the impact velocity vector and parallel to the impact plane. Each line of sight has constant $x$-$z$ values and varying $y$ values. The same set of $y$-values are used for each line of sight, and we integrate into the structure in the direction that $y$ is increasing (becoming less negative). Sample points are equally spaced along each individual axis. We then account for areas of significant opacity by adding additional steps ($y$ values) and interpolation points ($x$-$z$ pairs) as follows:

\begin{enumerate}
    \item We extend the lines of sight in both the positive and negative $y$ directions, still ensuring that all sample points are equally spaced. Extra line of sight padding is added until the first integration step along each line results in $\tau \approx \Delta r \ \alpha(r_1) < 1$.
    \item We pad the $x$ values and $z$ values (individually) until the lines of sight with the outermost $x$ and/or $z$ values are all optically thin (for photons traversing the entire line of sight).
\end{enumerate}

We enforce that any grid point outside of the (spherical) information region contributes no opacity. Test integrations which integrated over the whole information region were conducted and yielded near-identical results to this procedure. Not only is our partial integration faster, it allows us to integrate over smaller steps, thus resolving the features of photic surfaces in more detail.

We integrate along each line of sight until the point at which we reach an optical depth of 1 or above. If the integration reaches the end of the given line of sight before an optical depth of 1 is reached, we consider that line of sight to be transparent. We use these results to map the location and thermal structure of photic surfaces. For an optically thick line of sight $i$, at the point at which we reach $\tau \geq 1$, we record the temperature (averaged by material type):
\begin{equation}
    T_{i}^{\rm ps} = f_{\rm fe}  T_{\rm fe} + f_{\rm fors} T_{\rm fors}
\end{equation}

With our photosphere being represented as $n$ grid cells of equal area with temperatures ${T_{i}^{\rm ps}}$, we use the Stefan-Boltzmann law to calculate the flux of the photic surface (as viewed at a given angle):
\begin{equation}
    F = \frac{\sigma}{n \pi} \sum_{i = 1}^{n} ({{T_i^{\rm ps}}})^4,
\end{equation}
where $\sigma$ is the Stefan-Boltzmann constant. The factor of $\pi$ accounts for the fact that our photosphere, approximated as a black body, is a Lambertian surface. We calculate an effective temperature corresponding to this flux using:
\begin{equation}
   F = \frac{\sigma}{ \pi} (T_{\rm eff})^{4} \implies T_{\rm eff} = \bigg(\frac{F \pi}{\sigma}\bigg)^{\frac{1}{4}}. \\
\end{equation}
Finally, for a total emitting area of $A_{\rm tot}$, we calculate photospheric luminosity as:
\begin{gather}
    L_{\rm ps} = F A_{\rm tot}.
\end{gather}

Note that this luminosity does not correspond to the total radiating power of the impact structure, but rather the radiating power of what would be seen from a single viewing angle. As such, the luminosities calculated in this study are lower limits on the total luminosities of impact systems.

\section{Results and Analysis}
\label{sect:analysis}
In this section, we discuss the application of the pipeline presented in Section \ref{sect:methods} to the suite of impact simulations presented in Figure \ref{fig:impact_suite}. We first discuss the outcome of recalculating smoothing lengths and densities for the SPH particles in our simulations (Section~\ref{subsect:results:particles}). Then, we present photospheric maps in 2D and 3D after 24 hours of simulation time (Section \ref{subsect:photosphere_mapping}) and describe the evolution of photic surface effective temperatures and luminosities during the simulation (Section \ref{subsect:emission_and_evolution}). Throughout this section, we analyse the effect of initial choices of particle resolution, $N$, and maximum smoothing length, $h_{\rm max}$, on our results to assess the robustness of our method to the ``quality'' of the originally simulated SPH data (considering simulations with higher values of $N$ and $h_{\rm max}$ to be of better quality).

\subsection{Post-Processing of SPH Particles}
\label{subsect:results:particles}
As discussed in Section~\ref{subsect:data_processing:arbitrary}, it is not necessary to reassign the smoothing lengths and densities of individual SPH particles in order to calculate the photosphere of an impact structure.  However, doing so provides insight into the effect of the choice of $N$ and $h_{\rm max}$ on simulation output and thus the photospheres that we calculate. 

Figures~\ref{fig:reassign_rho}a~\&~\ref{fig:reassign_rho}c show histograms of the smoothing lengths and densities, respectively, of simulations with varying $h_{\rm max}$, but constant $N = 10^{6.5}$, before (dashed lines) and after (solid lines) post-processing. For many particles at $h_{\rm max}$, their recalculated smoothing length is orders of magnitude larger than $h_{\rm max}$, and consequently their recalculated density is many orders of magnitude lower than the minimum density, $\rho_{\rm min}$, imposed at runtime. For the values of $h_{\rm max}$ that we test, when $h_{\rm max} \geq 1 {\rm R_{\oplus}}$, the recalculated distributions of smoothing length and density approach convergence. However, when comparing these higher $h_{\rm max}$ cases to the $h_{\rm max} = 0.316 \ {\rm R_{\oplus}}$ case, there are significant differences in the frequencies of particles at certain smoothing lengths or densities between $0.316 \ {\rm R_{\oplus}} < h < 1 {\rm R_{\oplus}}$. This region is directly resolved at runtime by simulations with higher $h_{\rm max}$ values.

Figures~\ref{fig:reassign_rho}b~\&~\ref{fig:reassign_rho}d show equivalent histograms but for simulations with varying $N$ and fixed $h_{\rm max} = 1{\rm R}_{\oplus}$. In this case, the frequency is scaled so that all simulations are easily comparable in material density space. The recalculated particle density distributions essentially converge for $N \geq 10^{6}$, although simulations with more particles resolve lower density limits to the distribution. 

Together, these results suggest that for $h_{\rm max} \geq 1 {\rm R_{\oplus}}$ and $N \geq 10^{6}$ ($\rho_{\rm min}\lesssim10^{-2}~\text{kg m}^{-3}$ in this study), our SPH simulations were able to track the evolution of particle positions and smoothing lengths until pressure-gradient forces no longer had a substantial effect on the evolution of low-density regions of the impact structure. On the other hand, simulations with higher values of $\rho_{\rm min}$ were still evolving significantly under pressure-gradient forces when particles began to hit the density floor, and so the removal of pressure-gradient forces led to significant differences in trajectory and expansion of lower density material. Nevertheless, it is apparent that higher-$N$ simulations, when post-processed, are better at resolving the finer structure of the lowest density regions of the impact structure, which will prove important for the calculation of the photosphere (see Section \ref{subsect:photosphere_mapping}).

\subsection{Photosphere Mapping}
\label{subsect:photosphere_mapping}
\begin{figure*}
    \centering
    \includegraphics[width=\linewidth]{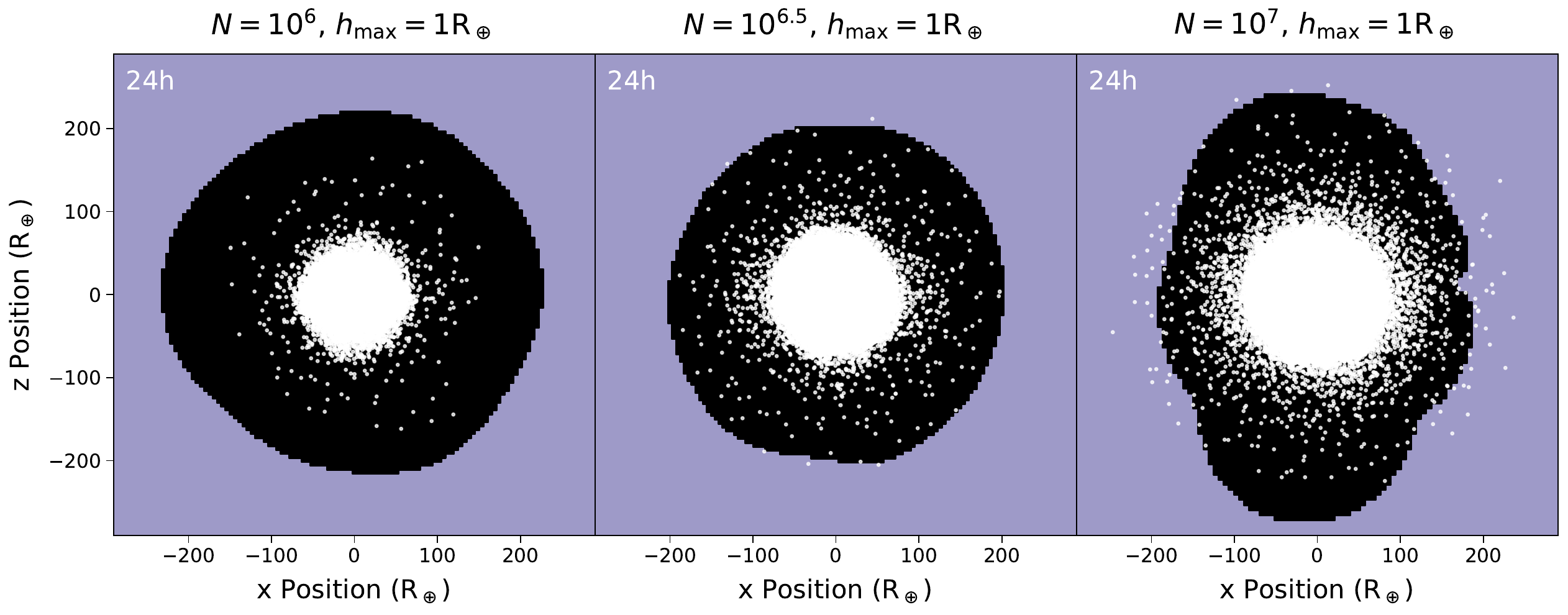}
    \vspace{-1em}
    \caption{Example photospheres for a given line of sight. Black cells represent lines of sight along which a travelling photon would certainly be blocked (and the structure is optically thick), whereas purple cells represent lines of sight along which travelling photons may not be blocked (optically thin). The central locations of SPH particles are plotted in white. All three maps show the result of the same collision, with numerical resolution ($N$) increasing to the right. The maps are plotted with the centre of mass of the impact structure at the origin, with the impact plane represented by the $x$-axis. For ease of comparison, each map is plotted for same ranges of $x$- and $z$-position values.}
    \label{fig:photic_surface_comparison}
\end{figure*}

Figure \ref{fig:photic_surface_comparison} shows three 2D maps of the photospheres of impact structures, indicating the difference between optically thin ($\tau < 1$, light purple) regions, and optically thick ($\tau \geq 1$, black) regions. The three maps correspond to simulations of the same impact scenario with an $h_{\rm max}$ of $1 \rm R_{\oplus}$, but with varying particle resolutions of the target ($N \in \{10^{6}$, $ 10^{6.5}$, $ 10^{7}\}$). The central locations of SPH particles in the simulation are also plotted for reference (in white). The photic surface for this example impact is very large (100s~$\rm R_{\oplus}$) with most of the ejecta being optically thick from this viewing angle.

Figure \ref{fig:photic_surface_comparison} gives an initial visual comparison of the sensitivity of our photosphere location method to the numerical resolution of simulation data. For $N = 10^{6}$, the optically thick region encloses all central locations of SPH particles, and the photic surface is defined by interpolation beyond the locations of the outermost SPH particles. However, for $N = 10^{7}$, there are several SPH particles locations that are outside of the optically thick region, and so the morphology of the photic surface is being defined by interpolation both within and beyond the particle location cloud. Indeed, the shape of the optically thick region becomes more complex with increasing resolution, as the lower mass particles can more finely describe the outermost regions of the expanding ejecta and therefore the edges of the photic surface. For example, the highest resolution ($N = 10^7$) presents some vertical aberration, being more elongated than its lower-resolution counterparts. However, despite these differences between different resolution simulations, the length scale and general shape of each optically thick region is encouragingly similar. In other words, despite particles being distributed further from the centre of the impact structure when $N$ is higher, we find that there is good consistency in the scale of the photic surfaces.

\begin{figure*}
    \centering
    \includegraphics[width=\linewidth]{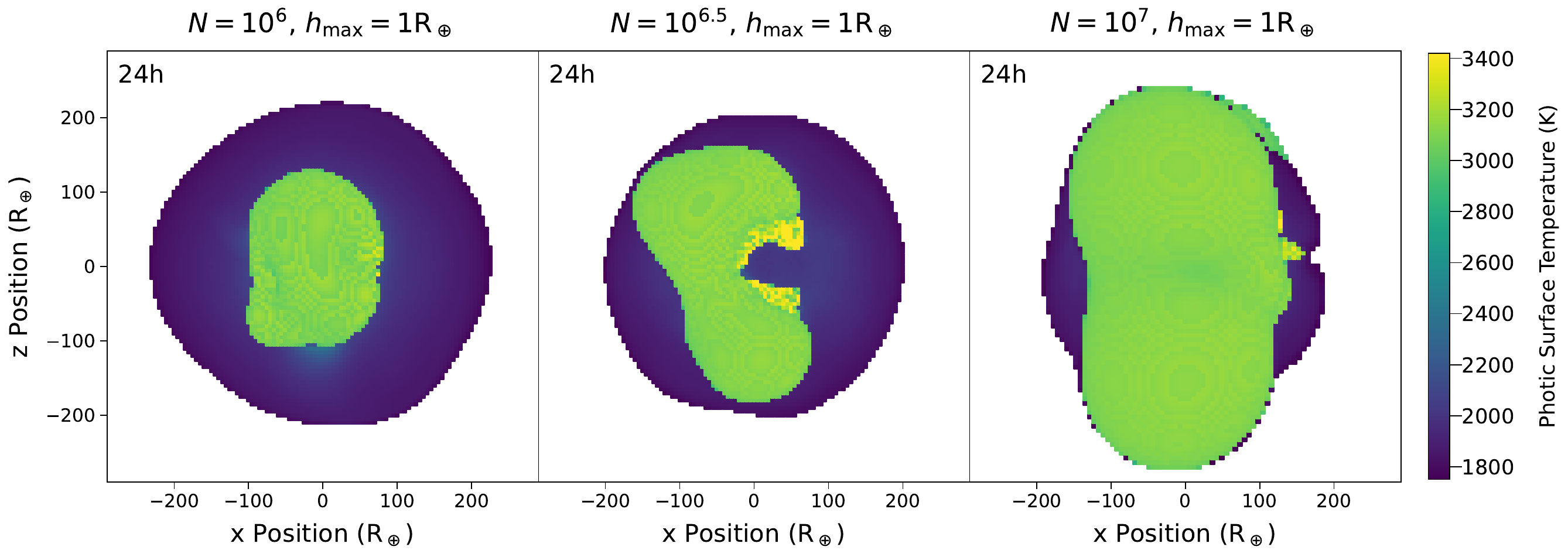}
    \vspace{-1em}
    \caption{Temperature maps of the surface $\tau \approx 1$ for example impact structures, derived from the same post-processed simulation data as the opacity segmentation maps shown in Figure \ref{fig:photic_surface_comparison}.}
    \label{fig:photic_surface_temp}
\end{figure*}

The temperature at the photic surface reveals a more complex structure and a greater dependence on simulation parameters. Figure~\ref{fig:photic_surface_temp} shows 2D photosphere temperature maps, for the same examples as in Figure~\ref{fig:photic_surface_comparison}. These plots show varied thermal structures, but all exhibit the same key features: a cool background that is somewhat circular from the line of sight we are considering; much hotter amorphous regions closer to the impact point (approximately the origin) which become larger with increased resolution; and some patches of even hotter material identified close to the impact plane ($x$-axis). 

\begin{figure*}
    \centering
    \includegraphics[width=\linewidth]{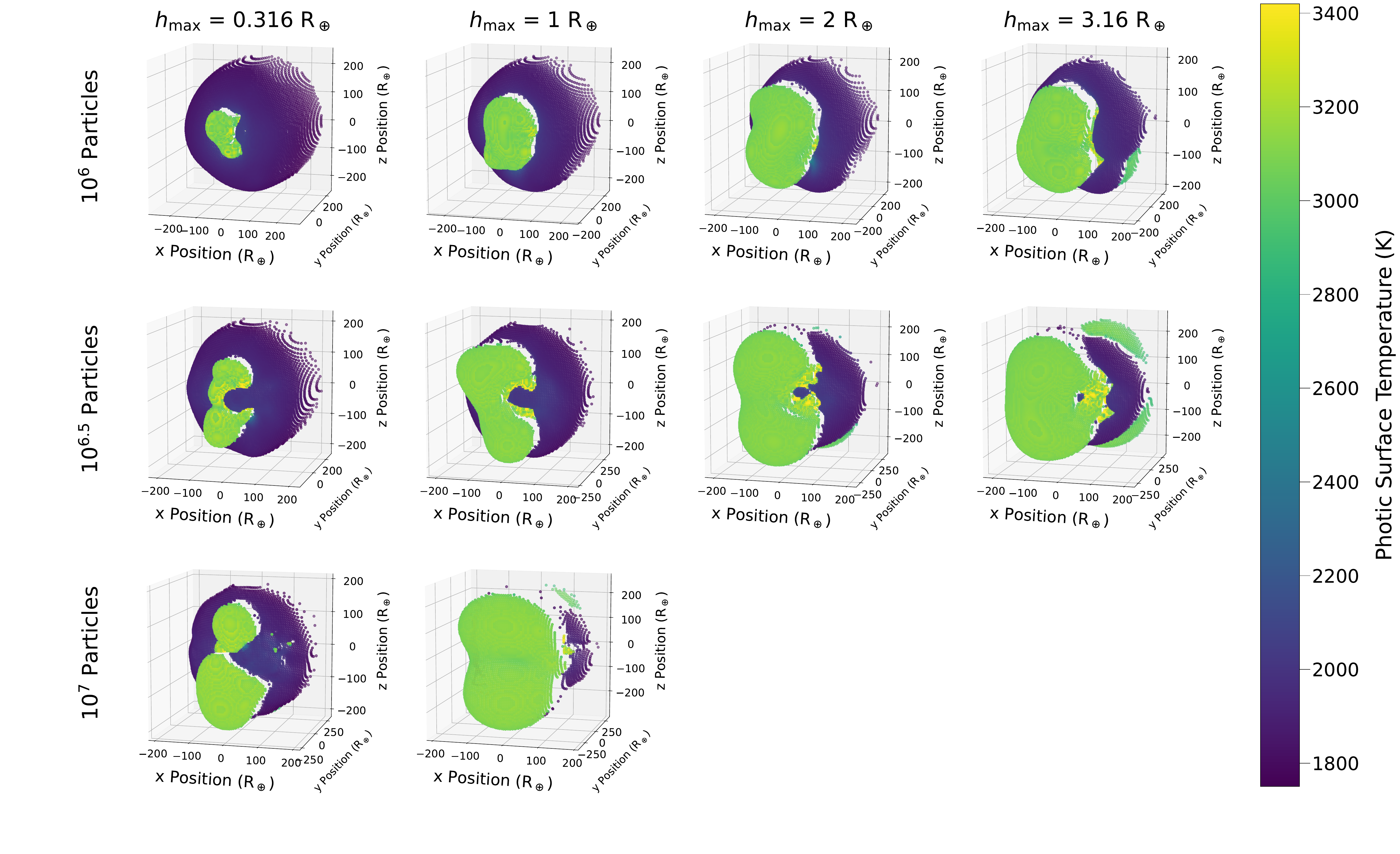}
    \caption{A grid of extracted photic surfaces resulting from the same giant impact scenario between two terrestrial bodies plotted in three dimensions, at a simulation time of 24\,h. Each row corresponds to the same approximate target resolution, and each column to the same maximum smoothing length, $h_{\rm max}$. The surfaces $\tau = 1$ are colour-coded by local temperature. Each line of sight has fixed $x$ and $z$ coordinates.}
    \label{fig:all_thermal_ps_3D}
\end{figure*}

These features can be better understood by looking at the photic surfaces in 3D, with the third dimension being the location at which the structure becomes optically thick (Figure \ref{fig:all_thermal_ps_3D}). 
Each photic surface features a (comparatively) cool, extended bulb where the structure becomes optically thick only over long distances. More centrally, the photosphere is defined by plumes of hotter material emanating from the impact point that cause the structure to be optically thick closer to the observer. The cooler material is shielding hot plumes on the opposite side of the structure from view, and in some cases, photons emitted from hot plumes on the other side of the structure can be seen from the viewing angle we have chosen (e.g., in the $N=10^{6.5}$, $h_{\rm max}=3.162 \rm \ R_{\oplus}$ case). Exposure of the hot plumes from the far side of the impact structure increases the area classified as optically thick in a 2D projection, and increases the luminosity of the impact (see Section \ref{subsect:emission_and_evolution}). The most extreme temperatures on the photic surface are from areas where we can effectively ``see through'' to the more superheated, central regions of the impact structure, closer to the point of impact. 

\begin{figure}
    \centering
    \includegraphics[width=\linewidth]{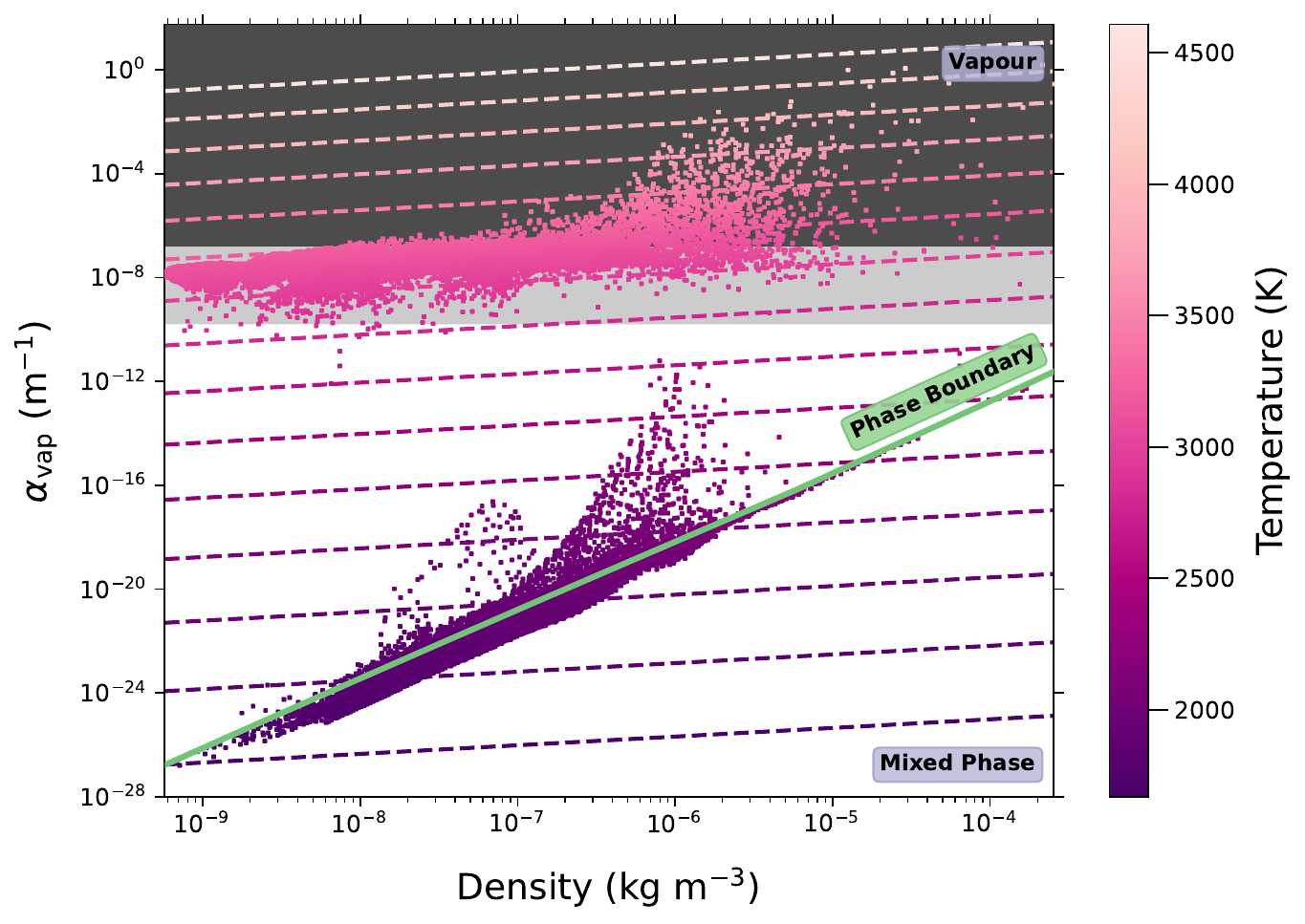}
    \caption{Visualising how the properties of vapour at all photic surface locations evaluated in this study (solid points) compare to the Kraus vapour absorptivity metric (Equation \ref{eqn:vapour_absorption}, from \protect\cite{kraus2012_vap}, shown as dashed lines). Absorptivity, $\alpha_{\rm{vap}}$, is plotted as a function of vapour density and colour-coded by vapour temperature. We use the same background colour-coding as Figure \ref{fig:absorptivity_metric_comparison} to indicate different optical thickness regimes. There is a clear temperature dichotomy between points on photic surfaces, with the lower temperature group in the white region requiring a significant contribution from condensate opacity for those regions of the photosphere to become optically thick. The higher temperature group is mostly contained within the grey and black region, hence hot vapour is playing the dominant role in optical thickness. This temperature dichotomy can be seen in Figure \ref{fig:all_thermal_ps_3D}. We also plot the phase boundary between pure vapour and mixed phase material for reference (solid green line), with pure vapour being above this line and mixed phase material below. Note that the photic surface points plotted below the phase boundary are actually considered to be on the phase boundary, but considering the order of magnitude of $\alpha_{\rm vap}$ in those regions, some systematic error is present.}
\label{fig:kraus_absorptivity_metric_with_photosphere_data}
\end{figure}

The phase of material varies across different regions of the photic surface. The hot plumes are almost entirely composed of vapour, with the absorption dominated by the opacity of vapour (Figure \ref{fig:cond_frac_all_ps} in Appendix \ref{appendix:photic_surface_condensate}). In the colder bulb, the material at the photosphere is a mixture of vapour and condensate. Here, the vapour is much cooler and less absorptive (Figure \ref{fig:kraus_absorptivity_metric_with_photosphere_data}), and therefore the opacity of these regions is dominated by the optical thickness of the condensate. Thus, a general pattern emerges: the photic surfaces we recover feature hot plumes containing significant portions of vapour, and a cooler background whose absorptivity is controlled by its condensate component.

The choice of user-specified simulation parameters can have a substantial effect on the thermal structure of the photosphere. As simulation particle resolution, $N$, increases, the hotter, vertical, vapour-dominated plumes become more extended, and the cooler background material displays more temperature variation. This is the natural consequence of analysing higher-quality data in which mass is not sequestered in larger particles, therefore providing a more accurate (likely more radially extended) mass distribution. Further, the simulations with higher resolution and higher $h_{\rm max}$ (i.e. lower density floors) are able to trace the pressure-driven expansion of the optically thick vapour plumes to lower densities, and so to later simulation times and greater spatial extents (see Section~\ref{subsect:results:particles}). The effect of resolution is particularly relevant to the hot plumes, as their properties are dictated by a relatively small number of particles near the photosphere, even for high resolutions. However, simulations with similar density floors produce vapour-dominated areas of photosphere that are of similar extents (simulations along right-leaning diagonals in Figure~\ref{fig:all_thermal_ps_3D} have similar minimum densities) suggesting that the primary control is how well the dynamics of the impact are captured in the simulation. In other words, lower initial $h_{\rm max}$ simulations can somewhat underestimate the expansion of (partially) vaporised debris (see Figure~\ref{fig:all_thermal_ps_3D}), which is an effect that cannot be remedied via a post-processing method based on particle position data.

Note that while restricting the radial mass distribution affects the size of vapour plumes, this does not translate to restricting the scale of the cooler bulb. In fact, we see a slightly larger condensate-dominated background for more restricted mass distributions resulting from lower $N$ values. When the particle distribution is more compacted, the information region will be smaller, but in all cases shown in Figure~\ref{fig:kraus_absorptivity_metric_with_photosphere_data}, the information region contains the same amount of mass. If the same amount of mass is being smoothed over a region that is (unrealistically) smaller, the result is that the density is higher, especially towards the edges of a structure that would have spread out more in better-resolved simulations. Furthermore, the particles that we are interpolating between have more mass for simulations with lower resolutions. Overestimations of local mass, and hence density, in the outermost zones of the information region can slightly increase the interpolated volume of condensate, which can become significant for an integration over such a large distance. Hence, at a given simulation time, the cooler background may be larger for more compact mass distributions. 

We have not presented photosphere maps for target particle resolution $N = 10^{5}$ because the structures were found to be optically thick for all lines of sight over the entire information region. Our method is therefore not guaranteed to accurately find the photosphere in this scenario. The minimum post-processed particle densities of simulations with $N = 10^5$ were $\sim 3.2 \times 10^{-6}$~kg~m$^{-3}$, and although this is lower than the density floors of simulations we can feasibly run, it still proved an insufficient ``smoothing out'' of mass for a photosphere calculation. The $10^{5}$ case can be contrasted with the $N=10^6$ case, where even though the projected transition between optically thick and optically thin (see Figure \ref{fig:photic_surface_comparison}) extends beyond projected particle locations, the transition is inside the information region (Section \ref{subsect:data_processing}). 

In summary, we can obtain relatively consistent sizes and temperatures of optically thick regions for a range of different resolutions and maximum smoothing lengths for $h_{\rm max} \geq 1 \rm R_{\oplus}$ and $N \geq 10^6$. A high enough resolution is required to have a sufficient number of SPH particles representing the relatively low-mass outer regions of the impact structure in order to capture the photosphere. To obtain a consistent morphology and/or temperature distribution, the combined effect of resolution and maximum smoothing length must also be such that the density floor (Equation \ref{eqn:rho_min_approx}) is low enough that the dynamics of the low density material defining the photosphere are tracked correctly until such a time that motion can truly be approximated as ballistic, and the lack of pressure forces in the later stages of the impact does not have a major effect on the calculated structure.

\subsection{Photosphere Emission and its Evolution Through an Impact}
\label{subsect:emission_and_evolution}

\begin{figure*}
    \centering
    \includegraphics[width=\textwidth]{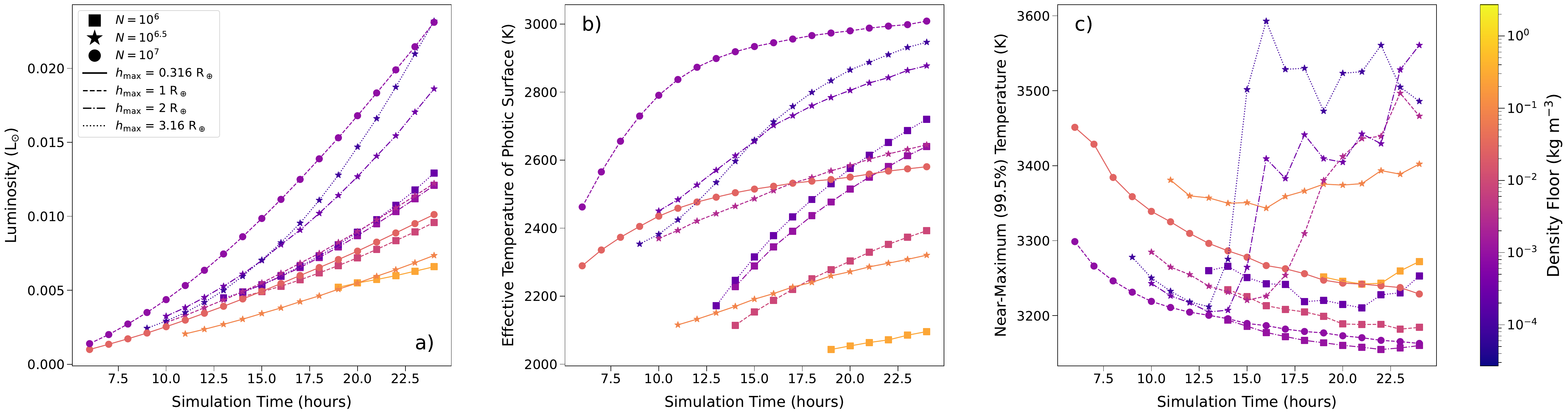}
    \caption{Evolution of the effective temperature and luminosity of a collisional system over time up to 24 hours. Resolution is indicated by symbol: for $N = 10^6$, data points are squares, $N = 10^{6.5}$ data points arestars and $N = 10^{7}$ data points are circles. Different $h_{\rm max}$ values are represented using different line styles, as detailed in the legends of each plot. The plot is colour-coded by simulation density floor. Plot a) shows photic surface luminosity evolution in units of solar luminosity, $L_{\odot}$, b) demonstrates the evolution of the effective temperature of the photic surface (black-body equivalent temperature of material ``that we see''), and c) shows the evolution of some of the highest temperatures recorded at the photic surface (with the 99.5th percentile used instead of the true maximum). }
    \label{fig:time_evolution}
\end{figure*}

Photospheric temperature maps, such as those shown in Figure \ref{fig:photic_surface_temp}, can be used to calculate the luminosity and effective temperature of impact structures. Figure \ref{fig:time_evolution} shows the properties of the photic surface (for a range of choices of $N$ and $h_{\rm max}$) as a function of time up to 24\,h. Note that we have not included results from the first few hours of the simulation, particularly for low $N$ simulations, as our method for calculating the opacity of the structure becomes computationally intensive if there are large volumes with very few particles (see Section~\ref{subsect:discussion:limitations_pp} for discussion). 

The luminosity of the impact structure increases steadily throughout the collision, reaching $\mathcal{O}(10^{-2}) L_{\odot}$ by 24~hrs. This luminosity increase is driven in part by the rapidly increasing size of the impact structure as unbound material travels out of the potential well of the colliding bodies, accelerated by the release of initially high-pressure, (partially-)vaporised material. Over time, hotter vapour plumes expand to obscure the material that contributes the opacity in the cooler ``background'' of the photosphere (see Section~\ref{subsect:photosphere_mapping} and Figure~\ref{fig:all_thermal_ps_3D}) which is ejected earlier in the collision. As a result, the effective temperature of the photosphere increases over time, reaching between 2000 and 3000~K within a few hours of the collision, further increasing the luminosity. However, the trend in effective temperature is concave as the vapour plumes, which have a more homogenous temperature, begin to dominate the burgeoning photic structure at later times. Therefore, although the increase in effective temperature contributes to the increase in luminosity with time, it is not the dominant driver at later times.

In general, the highest percentiles in temperature on the photosphere decreases with time (Figure~\ref{fig:time_evolution}c), but in some cases, such as those with $N=10^{6.5}$, a hotter part of the structure close to the impact plane becomes exposed later in time (Figure~\ref{fig:all_thermal_ps_3D}) leading to an increase in the near-maximum temperature. However, these small regions of hotter temperatures do not significantly affect the overall effective temperature or luminosity.

There is substantial variation in the calculated time evolution of luminosity and effective temperature, depending on the simulation data quality (Figure~\ref{fig:time_evolution}). For a fixed $h_{\rm max}$ (line style), luminosities are almost always higher for larger values of target resolution, $N$. This is not unexpected, since the motion and thermal properties of smaller amounts of mass can be modelled more accurately with more particles. Furthermore, for a given $N$ (marker style), the luminosity of the system is generally higher if $h_{\rm max}$ is larger, since $\rho_{\rm min}$ is smaller. More generally, we see an overall upwards trend for luminosity at 24\,h as $\rho_{\rm min}$ decreases, although dependence on $N$ tends to dominate.

We see similar trends in effective temperature (Figure \ref{fig:time_evolution}b). At a particular simulation time, for a given $h_{\rm max}$, effective temperature increases with $N$. For a given $N$, effective temperature is almost always larger when $h_{\rm max}$ is larger. In addition to these trends, we also see that for high enough $h_{\rm max}$, our results for a given $N$ begin to converge (e.g. see $N=10^6$ simulations marked by lines with squares in Figure~\ref{fig:time_evolution}). However, there are still significant variations between simulations with different $N$. This suggests that how well the mass distribution in the outer regions of the impact structure, and hence the morphology of the photic surface, is described becomes more significant than the premature neglect of pressure forces for sufficiently low minimum densities. This is consistent with the results of our reprocessing individual particles (Section~\ref{subsect:results:particles}), where we saw that, for the same $N$, the recovered density distributions of SPH particles were similar for sufficiently high vales of $h_{\rm max}$ (Figure~\ref{fig:reassign_rho}c), but that, for a given $h_{\rm max}$, increasing $N$ decreased the lowest density being resolved (Figure~\ref{fig:reassign_rho}d).

In some cases, a substantial portion of luminosity is actually due to emission seen from hotter material on the other side of the impact structure (see Figure \ref{fig:all_thermal_ps_3D}). For example, when $N = 10^{6.5}$ with $h_{\rm max} = 3.162 \ \rm R_{\oplus}$, this effect causes the luminosity at 24\,h to be larger than that of the case with $N = 10^7$ and $h_{\rm max} = 1 \rm R_{\oplus}$, even though the former has a lower average effective temperature, due to an increased photospheric area. (We do also see some emission from the other side of the structure for the $N = 10^7$, $h_{\rm max} = 1 \rm R_{\oplus}$ case.) 

With the aid of our post-processing pipeline, we are retrieving consistently large, heated, and luminous photic surfaces. However, although our post-processing method allows us to mitigate the effects of $N$ and $h_{\rm max}$ variation, our investigation is still limited by simulation quality and we have not reached parameters for which our results converge with both $N$ and $h_{\rm max}$. Based on the trends of luminosity and effective temperature with $N$ and $h_{\rm max}$, our results likely provide a lower limit on both $L$ and $T_{\rm eff}$, compared to a ``fully-resolved'' simulation with infinite particles and no $h_{\rm max}$. 

\section{Discussion}
\label{sect:discussion}

In this work we have presented a new pipeline for the analysis of giant impact simulation data to calculate the photosphere of impact structures, and applied this to an example giant impact to explore the sensitivity to simulation parameters. Our new pipeline includes the application of new tools/approaches to calculate material-wise properties at any location in the evolving structure, and the use of existing tools/data such as EoS and absorptivity models. The method we have presented begins to bridge the gap between theoretical models and direct observational signatures of giant impacts. 

In this section, we discuss how our results change our understanding of the detectability of terrestrial giant impacts, and the limitations of our post-processing methods.

\subsection{Feasibility of detection of giant impacts}
\label{subsect:feasibility_of_detection}

\begin{figure}
    \centering
    \includegraphics[width=\linewidth]{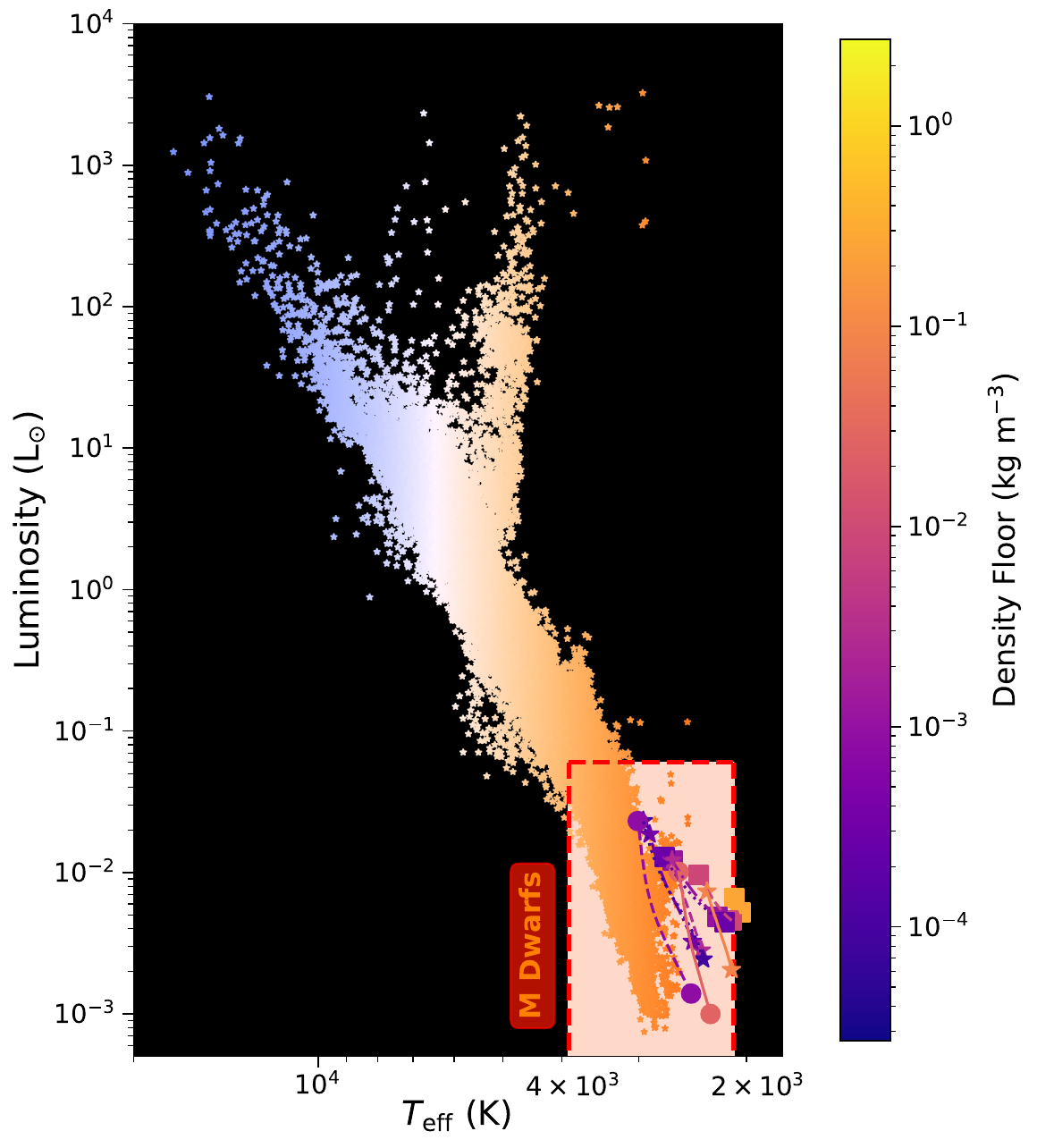}
    \caption{A Hertzsprung--Russell diagram of stellar temperature against luminosity of a selection of primarily main sequence stars, with the emission from terrestrial giant impacts superimposed. The evolution calculated from our giant impact simulations with each $N$ and $h_{\rm max}$ combination are shown by two identical symbols, corresponding to the first and last time for which we have calculated values for $T_{\rm eff}$ and $L$ (Figure~\ref{fig:time_evolution}), and dashed lines in the same colour as symbols show the evolution of photic surfaces in time through temperature and luminosity space. Choices of colour and marker are the same as Figure~\ref{fig:time_evolution}. Approximate bounds on the range of temperatures and luminosities that correspond to M Dwarf stars \protect\citep{mdwarf_review} are indicated by red dashed lines and shaded box. Our data points fall almost entirely in this range. Stellar data has been selected from Gaia Data Release 3, using the \textsc{Gaia Source} and \textsc{Astrophysical Parameters} tables \protect\citep{gaia-tables}. Colour indicates ``true'' stellar colour calculated from temperature, considering these stars to be perfect black bodies.}
    \label{fig:hr_diagram}
\end{figure}

Now that we have placed constraints on the temperature and luminosity of emission from an early-stage collisional system, we can assess whether the giant impact we have considered, and similar impacts between terrestrial bodies, would be astronomically observable. The direct thermal emission, at least within the first few hours after an impact, has a luminosity and an effective temperature comparable to those of stars. Indeed, if we compare the calculated $T_{\rm eff}$ and $L$ evolutions to stars using a Hertzsprung--Russell diagram (Figure \ref{fig:hr_diagram}), we see that the thermal emission is comparable to M Dwarf stars, for all combinations of $N$ or $h_{\rm max}$. This suggests that the feasibility of detecting the impact is comparable to that of detecting the sudden emergence of an M Dwarf companion around the host star of a planetary system. 

If we were to assume that the thermal emission produced by our terrestrial impact scenario at 24 hours can be approximated by black bodies with temperatures $T_{\rm eff}$, we find the peak wavelengths, $\lambda_{\rm peak}$, of the corresponding black body curves to be in the range of 0.963 -- 1.38 $\mu$m (varying slightly with simulation parameters, see Table \ref{tab:all_sim_data}). In other words, dominant emission wavelengths corresponding to the photic surfaces found in this study are near-infrared wavelengths (within the $Z$ to $J$ bands). Somewhat transient impact structures could therefore be detectable in data spanning visible and near-infrared wavelength ranges. In particular, new high-cadence wide-sky surveys such as the Legacy Survey of Space and Time (LSST) at the Vera C. Rubin Observatory may capture a large number of giant impacts, even if only in single observation windows.

However, our ability to quantify the probability of directly observing giant impacts is limited, as we cannot determine the long-term evolution and emission of the impact structure beyond the first days of an impact using an SPH impact code. This is because these codes do not currently include radiative cooling or decoupling of gas and condensates (see Section~\ref{subsect:discussion:limitations_SPH}). For context, the prolonged increase in infrared flux exhibited by ASASSN-21qj lasted $\sim 1000$ days. Further, the parts of impact structures that are observable in the days following an impact are unbound to the structure, and would be almost beyond the Hill sphere at 24\,h if the impact occurred at $\sim1$\,au around a Sun-like star. The continued evolution of the photosphere would be affected by interactions with the host star, and therefore become dependent on the orbit of the centre-of-mass frame of the collision and the mass/type of the host star \citep[e.g.][]{watt+2023,watt_planetary_2021}.

As unbound material continues to expand, it cools, condenses, and its optical depth increases. The unbound debris also ceases to be co-located with the bound post-impact body, as the range of ejection velocities, and hence orbits of the debris, shears the distribution around the host star \citep[e.g.][]{watt+2023,watt_planetary_2021}. Eventually, therefore, direct emission from the bound mass would become visible, at least from some viewing angles. Terrestrial post-impact bodies are themselves tens of times bigger than condensed planets and are expected to emit at approximately 2000--3000~K \citep{LockStewart17,lock2018_moon}, potentially giving them a substantial luminosity of their own \citep{KenworthyLock23}. The long-term emission of the impact structure is likely dominated by a combination of emission from the bound mass, and the cooled unbound material passively heated by the host star and, to some degree, by radiation from the bound post-impact body. Constraining the thermal evolution and lifetimes of post-impact bodies for longer than we can accurately simulate them in SPH will require the development of new analysis techniques. Further, the transition from the substantial luminosities within hours of the impact to the observation of direct emission from the bound post-impact body is likely to be complicated, yet key to both identifying the signatures of planetary impacts and the probability of detection.

We can also use our data to assess observation of another proposed signature of giant impacts: the transit of impact debris between the host star and Earth \citep[e.g.][]{KenworthyLock23}. To this end, we draw attention to the sheer scale of the photic surfaces shown in Figure \ref{fig:all_thermal_ps_3D}. In all cases, these photic surfaces span multiple hundreds of Earth radii within just 24\,h of the collision, and would certainly be large enough to entirely eclipse a Sun-like star. We therefore conclude that it is reasonable for a significant stellar dimming event to follow a giant impact at any time after the first couple of hours after impact. Of course, the possibility of seeing an associated transit would be highly dependent on the orbital dynamics of the star system in which the impact occurs, the properties of the system's host star, the orientation of the plane of collision to the line of sight between the star and Earth, and the duration over which the debris cloud remained optically thick. 

Finally, it is important to stress that this work contains analysis of a single impact scenario -- an oblique impact between two identical proto-Earth-like objects -- at one viewing angle. The assessment of the effect of different impact scenarios (e.g. different masses, impact velocities and impact angles) and viewing angles will be addressed in a future publication \citep{taylor_paper3}.

\subsection{Limitations of post-processing and photosphere calculations}
\label{subsect:discussion:limitations_pp}

While the development of our post-processing pipeline has opened up examination of the photic surfaces of impact structures, there are several important limitations that must be considered. Firstly, the computational efficiency of our analysis depends on the arrangement and number of particles in the SPH output. The sparser the particle distribution, the longer it takes to solve for proxy-smoothing lengths at arbitrary locations. The photosphere-finding procedure must begin by studying optical depth in the outer (information-poor) regions of the impact structure before moving inwards. For our purposes, this makes applying our post-processing pipeline to simulations with (relatively) small numbers of particles computationally expensive. In some cases, the particle arrangement is so compact that the transition from optically thin to optically thick material is located outside of the information region (e.g., simulations with $N = 10^5$ in our test suite). 

Secondly, we have so far only implemented basic post-hoc radiative transfer. We have used absorptivity models for the vapour/ super-critical fluid that average absorption over all wavelengths, ignored scattering and re-emission, and have treated emission from the photosphere as that of a black body. These assumptions preclude us from conducting a detailed comparison between our results and data from astronomical instruments \citep[e.g. as recorded by surveys such as ASASSN or WISE, used by][]{KenworthyLock23}. However, at this stage, there are two major barriers that need to be overcome before a more sophisticated radiative transfer calculation would even be worthwhile: we cannot include sophisticated models of chemical equilibrium, nor assess absorptivity at different wavelengths. In our analysis, we treat the silicate and iron as separate, non-interacting materials, and study the phase of each with separate EoS after enforcing pressure equilibrium. In reality, the metal and silicate would equilibrate and produce a system of multiple species, whose speciation could change dramatically across the system, with dominant species varying from SiO$_2$ to O, and MgO to H \citep{lock2018_moon,Fegley+23}. Previously, physical chemistry, Gibbs-free energy minimization models have been applied to the outer regions of impact structures \citep{lock2018_moon,Fegley+23}, but this would add significant computational expense to each integration step. Further, we assume that material outside of the photosphere is dominated by silicate, and so the silicate-iron vapour mix has the absorptivity of a pure silica vapour. Calculating wavelength-dependent absorption from multiple species would add significant computational load (particularly when including processes such as scattering and emission), but the more pressing issue is the lack of wide-ranging absorption models for some species. For example, the authors are not aware of a comprehensive high-temperature gas-phase FeO opacity model, but FeO would be a significant component of a silicate vapour, particularly if enriched with metal. Further, the effect on absorption from interactions between molecules (e.g., collision-induced absorption and pressure-broadening) is unknown. Further work is required to produce chemistry-informed wavelength-dependent models for calculating emission from impact structures.

Despite these limitations, many of the tools we have developed are highly robust within the framework of SPH and could be usefully applied elsewhere. For example, we have demonstrated that re-calculating smoothing lengths with smoothing length limits (i.e. $h_{\rm max}$) removed in a way that is consistent with the mathematical formulation of SPH is a practical and very useful form of post-processing that can retrieve approximate density and thermodynamic properties of material that has hit the density floor. Care should be taken while using this approach, however, as material that approaches the density floor during runtime will still not have had its motion and evolution calculated including pressure-gradient forces. This method is therefore best applied to material that is following close to ballistic trajectories at the point it hits the density floor, such as in the case of ejecta, low-mass circumplanetary disks, or other low-density systems.

We have also shown the potential of a non-particle centric interpolation technique, by which a proxy smoothing length is calculated at an arbitrary location  to interpolate density in a way that is consistent with the key mathematical principles of SPH, and solve for the thermodynamic properties (including phase) of material. Although the SPH implementations used for impact simulations do not allow material mixing, our pseudo-particle technique can separate out properties interpolated from different materials into the contribution from each material, allowing calculation of properties of individual components and the mixture. In future, this approach could be used to post-hoc calculate chemical equilibria between materials (as discussed above) and could even be adapted to model true material mixing and thermodynamic equilibration during runtime.    

\subsection{Limitations of current SPH techniques for determining observability}
\label{subsect:discussion:limitations_SPH}

Despite our best efforts, the primary limitation to the calculation of the emission of post-impact bodies from impact simulations is still the quality of the simulation itself. The calculated emitting areas, luminosities and effective temperatures of the photosphere that we obtain are relatively consistent between simulations with varying resolution, $N$, and maximum smoothing length, $h_{\rm max}$, but it would be incorrect to say that these results have reached convergence. For low enough minimum densities ($\rho_{\rm min}\lesssim10^{-3}~\text{kg m}^{-3}$), the distribution of the recalculated densities of individual SPH particles converges (Figure~\ref{fig:reassign_rho}), showing that the expansion of debris is tracked until sufficiently late in evolution that the material has become mostly ballistic. However, the low-density tail extends to lower densities for higher resolution simulations, and it is this low density material that controls the detailed topography of the photosphere. Still, for simulations with $N\geq 10^{6.5}$, the amount of material in the low density tail becomes more consistent and the topography, luminosity and temperature of the photosphere are similar. Based on this and other considerations (such as feasibility of interpolation), we recommend that our post-processing pipeline is applied to impact simulations where $N \geq 10^{6.5}$ and $h_{\rm max} \geq 1 \rm R_{\oplus}$, to ensure the results suffer less from resolution dependence but are simultaneously computationally tractable. Note that this recommendation may only apply to similar impact scenarios.

In addition to the issues with particle number and minimum densities, there are more fundamental limitations of the current implementations of SPH used in impact studies that restrict our ability to determine the emission and evolution of colliding systems. One of the most constraining of these is that current impact codes do not have the ability to include radiative transfer at run time. We have shown here that most of the impact structure is optically thick and so the lack of radiative transfer will not have a significant effect on impact outcomes that are controlled by the dynamics of the inner part of the structure. However, for the outermost regions of the structure, the transfer and loss of energy by radiation could be much more significant. Given the low density of the near-photosphere regions and the high luminosity of the system, the energy lost by radiation from the outer portion of the system could substantially cool these regions. In particular, the hot vapour plumes, which are the most luminous parts of the impact structure, would lose a significant fraction of their internal energy and partially condense. Neglect of radiative transfer means the calculated system becomes increasingly inaccurate the longer the simulation time, and is another reason that we do not simulate beyond 24\,h.

How radiation would affect the evolution of the structure is uncertain. One effect would be to reduce the vapour pressure of the structure, retarding the acceleration of the vaporised or partially-vaporised material at the edge of the structure (at and beyond the photosphere). However, optically thick material behind the leading edge would still be expanding (now with a lower resistance from vapour pressure in the material in front of it) and could still drive the optically thin material outwards while itself still within the photosphere. Depending on the coupling of condensates and vapour, vaporised material below the photosphere might alternatively sweep past the photosphere condensates and start to lose energy by radiation in its turn. Heat would be supplied to the photosphere from below which would also slow its cooling. Determining the overall impact of such processes is not possible without directly including radiative transfer and multiphase physics in impact codes. However, it is likely that the photosphere would be somewhat smaller, due to the reduced driver for material expansion, and cooler, due to more of the photosphere being on the liquid-vapour phase boundary, than we have calculated.

\begin{figure*}
    \centering
    \includegraphics[width=0.9\linewidth]{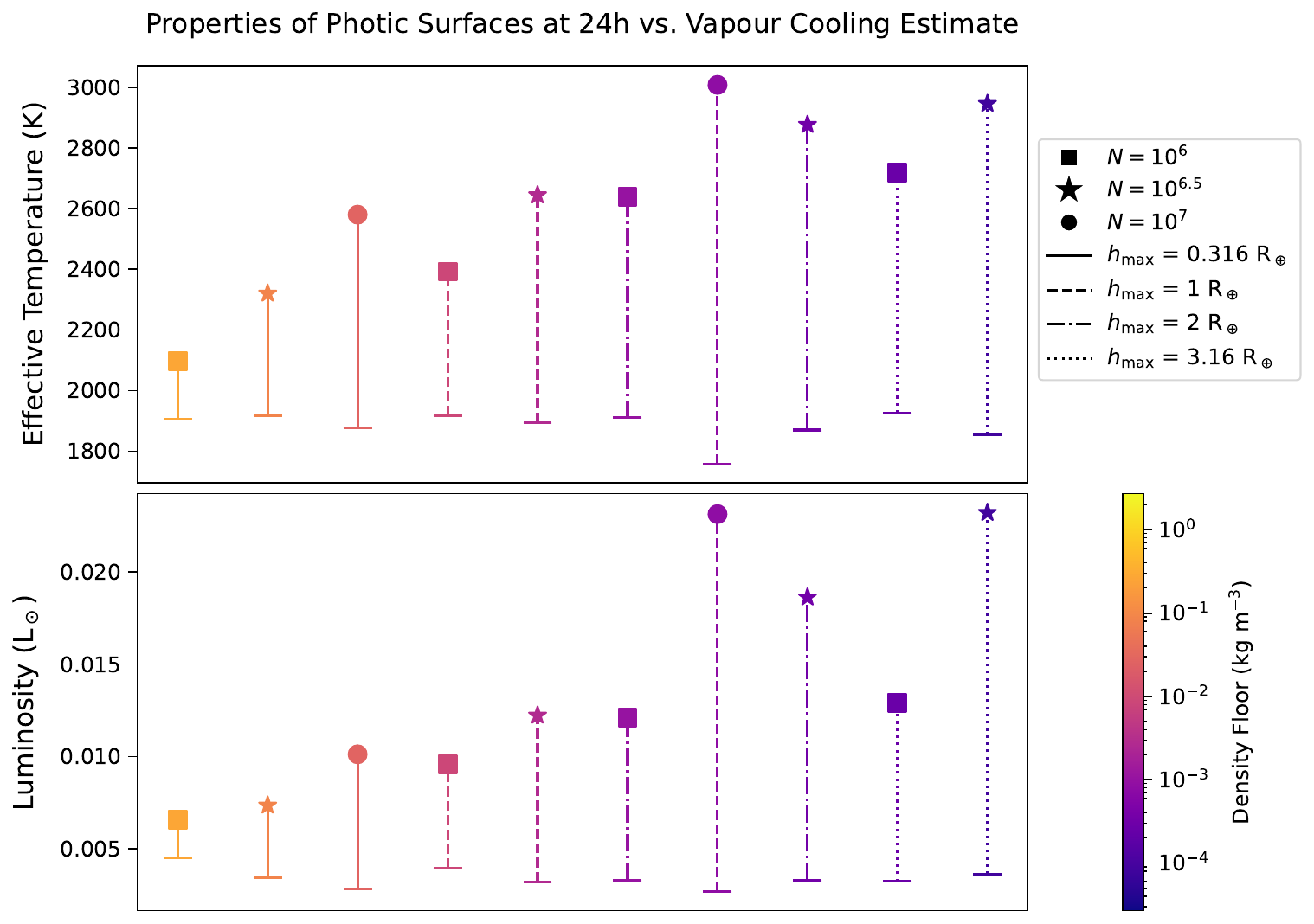}
    \caption{Condensation of the pure-vapour regions of the photosphere would reduce the effective temperature and luminosity of impact structures, by up to a factor of a few. Markers (squares, stars and circles) represent the effective temperature (top) or luminosity (bottom) of our extracted photospheres at 24\,h. They are plotted with a lower bound (horizontal bar) that represents the value of $T_{\rm eff}$ or $L$ when the pure vapour regions of the photospheres are isobarically cooled until they reach the vapour-condensate phase boundary. Markers and line styles indicate differences in $N$ and $h_{\rm max}$, consistent with Figure~\ref{fig:time_evolution}, and results are ordered by decreasing density floor.}
    \label{fig:tracking_back}
\end{figure*}

To approximate the effect of condensing vapour-dominated areas of the photosphere on the overall emission from the impact structure, we repeat our analysis but assuming that all pure vapour regions on the photosphere are in fact at the temperature of the liquid-vapour phase boundary at the same pressure (see Figure \ref{fig:tracking_back}). As expected, condensation of the photosphere significantly reduces the luminosity and effective temperature of the colliding system. However, the luminosities of these objects are still comparable to M Dwarfs (Section~\ref{subsect:feasibility_of_detection}), even if their effective temperatures are slightly too cold (less than 2100 K).

Another key set of processes that is currently lacking in SPH impact models is multiphase physics. If an SPH particle is at a thermodynamic state such that it would be a mixture of phases (most relevant in our case, vapour and solid/liquid condensate) then it is assumed that those phases are entirely coupled and in thermal equilibrium. Our post-processing scheme, by way of looking at each time snapshot in isolation, makes the same assumption. In reality, there is potential for decoupling of the condensate and vapour, as the vapour may be accelerated by pressure gradient forces that do not directly have much influence on the condensates. Such processes could increase the vapour fraction but reduce the total density of material in the outer regions of the structure, potentially resulting in a hotter photosphere, but this does render the material at the photosphere more susceptible to condensation (see above). The magnitude and nature of the effect of vapour-condensate decoupling on the luminosity of the impact structure is therefore uncertain.

There is also significant uncertainty on the size-distribution of condensates near the photosphere. Throughout this study, we have made the assumption that the diameter of condensate droplets is $D_0 = 1 \rm \, cm$, and constant throughout the entirety of the structure (see Section \ref{subsect:ot_theory} for justification). In reality, the droplets would be in a size distribution which varies throughout the structure, controlled by processes such as: nucleation from decompressing vapour; break-up of melt-dominated systems by instabilities or bubble nucleation; break-up due to gas drag; and potential cohesion/break-up due to mutual collisions. A different value of $D_0$ (or a non-uniform size distribution) would alter the observed geometry of the photic surface. Decreasing $D_0$ would render a given volume of condensate more absorptive, thereby expanding the radius of the cold, condensate background seen of the photic surfaces in Figure \ref{fig:all_thermal_ps_3D}. This radial expansion would result in a more ``engulfing'' cold condensate shield, blocking larger portions of the hot vapour plumes so that they effectively appeared smaller. This expansion of colder material would decrease the effective temperature (increasing $\lambda_{\rm peak}$), but may result in an increase in photic surface luminosity, since the projected area of the photic surface would be larger. For a decrease of $D_0$ by an order of magnitude, we find photospheres with similar luminosities but lower effective temperatures. However, from a computational standpoint, the amount that we can decrease $D_0$ is restricted by the fact that we do not want to interpolate beyond the information region. The smaller our $D_0$ value, the further the transition between optically thick and optically thin material is from the centre of mass of the post-impact structure. Significant advancements in our understanding of multiphase physics during giant impacts is required to accurately determine the size distribution of droplets and so their effect on emission.

\section{Conclusions}
\label{sect:conclusions}
In this paper, we begin to explore the feasibility of detecting collisions between terrestrial planets in other star systems based on their direct thermal emission \citep[as proposed recently for ice-giant planets by][]{KenworthyLock23}. To achieve this, we developed a novel post-processing pipeline and applied it to data from SPH simulations of an example giant impact scenario. Our pipeline emulates the realistic material and thermal data needed to relate simulations to observations, information which is not available from raw simulation data due to compromises made for the sake of computational efficiency.

We applied our pipeline to a single terrestrial impact scenario across a range of target resolutions and maximum smoothing lengths, and extracted 2D and 3D maps of the location of thermal emission (photic surfaces) in each case. These maps were used to calculate the luminosity and effective temperature of thermal emission after 24\,h of simulation time. (The lack of radiative transfer and multiphase physics in SPH impact simulations means we cannot reliably apply our results beyond a day or so after the collision.) For all resolutions and maximum smoothing lengths, these temperatures and luminosities were shown to be comparable to M Dwarf stars, implying that terrestrial impact structures are, at least transiently, astronomically observable.

All photic surfaces exhibited peak wavelengths in the near-infrared (when assumed to be black bodies), implying that high-cadence telescopes observing visible and near-infrared wavelength ranges would be best suited to detecting the impact structures that form after terrestrial giant impacts. As such, our work suggests that giant impacts may be frequently observed by new high-cadence wide-sky surveys such as the Legacy Survey of Space and Time (LSST) at the Vera C. Rubin Observatory. 
\\

\noindent\textbf{Key takeaways about SPH methods and our post-processing pipeline:}
\begin{enumerate}
    \item Through use of our post-processing pipeline, it is possible to effectively emulate density and other thermodynamic properties of material that raw data does not capture due to artificial limits on SPH particle size. In this use case, the pipeline allows us to recover the properties needed to accurately locate an optical depth of $\tau = 1$ in an impact structure. We can identify the photic surface even in very low density regions of these structures.
    \item We have demonstrated that interpolating smoothing lengths without smoothing length limits (i.e. no $h_{\rm max}$) in a way that is consistent with the mathematical formulation of SPH is a useful form of post-processing.
    \item We have highlighted the potential of previously unexplored interpolation techniques, e.g. generating a proxy smoothing length at an arbitrary location that is not the centre of an SPH particle.
    \item SPH does not model materials mixing between particles. We have demonstrated how to separate out properties interpolated from different materials into the contribution from each material, at any arbitrary location.
    \item This work makes a general case for ensuring awareness of the limitations of simulations and theoretical models. These limitations may be essential to obtaining any simulation output at all, but will likely affect data and preclude further analysis if left unprocessed. In the context of SPH simulations, this means ensuring that simulations are run with parameters conducive to resolving the features of interest, at least with the aid of post-processing.
\end{enumerate}

\noindent\textbf{Key takeaways about the detectability of impacts from the application of our post-processing pipeline:}
\begin{enumerate}
    \item Our results suggest that oblique collisions between identical proto-Earth-like bodies produce emission at luminosities and temperatures comparable to M Dwarf stars at $\sim$~1 day after impact, with peak wavelengths in the near-infrared ($Z$ to $J$ bands). These impact structures may therefore be astronomically observable and could be studied by upcoming space missions such as LSST. Future work will explore terrestrial giant impact observables for a wider range of impact parameters and viewing angles \citep{taylor_paper3}.
    \item The photic surfaces of our example impact structures have distinctive features controlled by different phases of material: hotter ($\sim3000$~K), entirely vapour plumes whose opacity is dictated by that of the vapour; and a cooler ($\sim2000$~K) background whose opacity is governed by a significant condensate mass fraction.
    \item The detailed morphology of photic surfaces calculated from simulations with different target resolutions, $N$, and maximum smoothing lengths, $h_{\rm max}$, can vary significantly, but the effective temperature and luminosity values remain reasonably consistent in all tested cases. However, impact structures are extremely dynamic, and our models of them are limited by simulation quality. To ensure a broadly similar morphology we recommend that studies interested in the lowest-density outer regions of similar impact structures conduct simulations with $N \geq 10^{6.5}$, and $h_{\rm max} \geq 1 \rm R_{\oplus}$ (or $\rho_{\rm min}\lesssim10^{-3}~\text{kg m}^{-3}$ for different implementations of SPH).
    \item The photic surfaces extracted in this study span hundreds of Earth radii. Thus, optically-thick debris is capable of entirely eclipsing stars within several hours of an impact, and thus producing detectable stellar dimming events should this debris pass between a host star and Earth.
    \item Our results are consistent with observations of what has been suggested to be direct emission and subsequent stellar dimming following a giant impact between two ice giants, as reported by \cite{KenworthyLock23}.
\end{enumerate}

%% Please use the acknowledgment and contribution environments. This will 
%% be anonomyized when the "anonymous" style option is used. 
\begin{acknowledgments}
LT acknowledges joint funding from UKRI STFC (grant No. ST/X508263/1) and the University of Bristol (School of Physics
HH Potter PhD Scholarship Fund). SJL acknowledges support from NERC, UK (grant No. NE/V014129/1). PJC and ZML acknowledge financial support from the UK Science and
Technology Facilities Council (grant No. ST/V000454/1 and
ST/Y002024/1)

The research in this paper made use of the \textsc{swift} open-source simulation code (\url{http://www.swiftsim.com}, \cite{2018swift}) version 0.9.0. This work used the computational facilities of the Advanced Computing Research Centre, University of Bristol - \url{http://www.bristol.ac.uk/acrc/}. The authors acknowledge the use of resources provided by the Isambard 3 Tier-2 HPC Facility. Isambard 3 is hosted by the University of Bristol and operated by the GW4 Alliance (https://gw4.ac.uk) and is funded by UK Research and Innovation; and the Engineering and Physical Sciences Research Council [EP/X039137/1]. This work has made use of data from the European Space Agency (ESA) mission Gaia (\url{https://www.cosmos.esa.int/gaia}), processed by the Gaia Data Processing and Analysis Consortium (DPAC, \url{https://www.cosmos.esa.int/web/gaia/dpac/consortium}). Funding for the DPAC has been provided by national institutions, in particular the institutions participating in the Gaia Multilateral Agreement. Generative artificial intelligence (AI) was not used to conduct this research or to produce this paper.

This work was carried out using the open source \textsc{swift} SPH code, which is available through GitHub (XXXX). The exact version in this work has been archived on Zenodo. The newly developed post-processing pipeline, input files and final output files for all SPH simulations, and scripts to reproduce all figures in this work will be made available through GitHub and archived on Zenodo upon acceptance of the manuscript.
\end{acknowledgments}

\begin{contribution}
%%This section gives authors the space to recognize author contributions. The text inside this environment is NOT counted towards the total word quanta. At a minimum, manuscripts are expected to include this text:

SJL and LT came up with the initial research concept. LT ran the simulations, developed the post-processing method, and carried out the analysis, aided by the other authors. LT was responsible for drafting, collating and submitting the manuscript. SJL, PJC and ZML edited the manuscript.

%% But authors are expected to provide more specific details, e.g. 
%%
%%SC was responsible for writing and submitting the manuscript.
%%WWM came up with the initial research concept and edited the manuscript.
%%OTS obtained the funding and edited the manuscript.
%%EBF provided the formal analysis and validation. He also edited the manuscript.
%%GEH Supervised the undergraduates, wrote the software and administers the project github and Zenodo repositories.
%%
%% Authors can use the Contributor Role Taxonomy (CRediT) at
%% for ideas on how write a good statement tailored to their needs.

\end{contribution}

% \facilities{HST(STIS), Swift(XRT and UVOT), AAVSO, CTIO:1.3m, CTIO:1.5m, CXO}

\software{\texttt{SWIFT} \citep{2018swift},  
          \texttt{SEAGen} \citep{kegerreis+19_seagen}, 
          \texttt{WoMa} \citep{Ruiz_Bonilla_2020},
          \texttt{NumPy} \citep{numpy},
          \texttt{SciPy} \citep{scipy},
          \texttt{matplotlib} \citep{matplotlib}
          }


\appendix
\section{Simulation Parameters, Settings, and Results}
\label{appendix:sim_data_table}
Key parameters and results of all \textsc{swift} simulations conducted as part of this study are presented in Table~\ref{tab:all_sim_data}.

Each simulation used a $b_4$ cubic spline kernel, where
\begin{equation}
    b_4(r) = {\rm max} \{(1 - r)^3, 0\} - 4 \ {\rm max} \{({\small\frac{1}{2}} - r)^3, 0\},
\end{equation}
for $ r \geq 0 $. In three dimensions, this function is normalised by the constant $C = \frac{16}{\pi}$, such that when the normalised kernel function is integrated over the unit sphere, $S$, we find that:
\begin{equation}
    C \int_{S} b_4(r) \ dS = 1.
\end{equation}

Such a normalisation condition is essential to the conservation of total simulation mass \citep{Price_2012} because it ensures that when the kernel, $W$, is integrated over a smoothing sphere, $V$, with kernel support radius $H$, we find that
\begin{equation}
    \int_{V} W( \mathbf{x}) \ dV = 1,
\end{equation}
since
\begin{equation}
    \begin{aligned}
        \int_{V} W( \mathbf{x}) \ dV &=
        \int_{S} H^3 \ W( |\mathbf{x}|/H) \ dS \\ &=
        \int_{S} w(r) \ dS \\ &=
        C \int_{S} b_4(r) \ dS \; .
    \end{aligned}
\end{equation}

%  and normalised over the unit sphere
% \begin{equation}
% \int_{|q'|\leq1} w(|q'|)\mathrm{d}^3 q'= 1 \; ,
% \end{equation}
% and $q'$ is an integration variable. 

The $b_4$ kernel has a value of $\gamma = \frac{H}{h} = 1.825742$. Furthermore, as described in Section~\ref{sect:sph_intro}, all of our \textsc{swift} simulations use \texttt{resolution\_eta~$=~1.2348$} when iteratively solving for smoothing lengths, which corresponds to a neighbour number approximation of 48.

\begin{table*}
    \centering
    \begin{tabular}{ccccccccc} % eight columns, alignment for each
        \hline \vspace{-1em} \\
	{$N$} & $h_{\rm max}$ ($\rm R_{\oplus}$) & $\rho_{\rm min}$ ($\rm{kg \ m^{-3}}$) & $R_t = R_i$ ($\rm R_{\oplus}$) & $M_t = M_i$ ($\rm M_{ \oplus}$) & $v_c = v_{\rm esc}$ (${\rm km \ s^{-1}}$) & $T_{\rm eff}$ at 24\,h (K) & $L$ at 24\,h (0.01 $L_{\odot}$) & $\lambda_{\rm peak}$ ($\mu$m)  \vspace{0.25em}\\
    
    \hline \vspace{-0.75em}
    \\
    $10^5$ & 1 & $8.58 \times 10^{-2}$ & 0.970 & 0.897 & 10.757 & - & - & -  \\
    $10^5$ & 2 & $1.07 \times 10^{-2}$ & 0.970 & 0.897 & 10.757 & - & - & -  \\
    $10^5$ & 3.162 & $2.71 \times 10^{-3}$ & 0.970 & 0.897 & 10.757 & - & - & - \\
    
    % \hline
    \vspace{-0.5em} \\ 
	$10^6$ & 0.316 & $2.73 \times 10^{-1}$ & 0.983 & 0.897 & 10.685 & 2100 & $0.659$ & 1.38 \\
    $10^6$ & 1 & $8.62 \times 10^{-3}$ & 0.983 & 0.897 & 10.685 & 2390 & $0.959$ & 1.21 \\
    $10^6$ & 2 & $1.08 \times 10^{-3}$ & 0.983 & 0.897 & 10.685 & 2640 & $1.21$ & 1.10  \\
    $10^6$ & 3.162 & $2.73 \times 10^{-4}$ & 0.983 & 0.897 & 10.685 & 2720 & $1.29$ & 1.07  \\
    
    % \hline
    \vspace{-0.5em} \\
    $10^{6.5}$ & 0.316 & $8.66 \times 10^{-2}$ & 0.996 & 0.897 & 10.617 & 2320 & $0.736$ & 1.25 \\
    $10^{6.5}$ & 1 & $2.74 \times 10^{-3}$ & 0.996 & 0.897 & 10.617 & 2650 & $1.22$ & 1.10 \\
    $10^{6.5}$ & 2 & $3.42 \times 10^{-3}$ & 0.996 & 0.897 & 10.617 & 2880 & $1.86$ & 1.01 \\
    $10^{6.5}$ & 3.162 & $8.66 \times 10^{-5}$ & 0.996 & 0.897 & 10.617 & 2950 & $2.32$ & 0.983 \\

    % \hline
    \vspace{-0.5em} \\
    $10^{7}$ & 0.316 & $2.69 \times 10^{-2}$ & 1.000 & 0.897 & 10.596 & 2580 & $1.01$ & 1.12 \\
    $10^{7}$ & 1 & $8.51 \times 10^{-4}$ & 1.000 & 0.897 & 10.596 & 3010 & $2.31$ & 0.963 \\
    
	\vspace{-0.75em}
    
    \\ \hline \hline
    \end{tabular}
    \caption{Effective photic surface emitting temperature ($T_{\rm eff}$), luminosity ($L$) and peak wavelength ($\lambda_{\rm peak}$) calculations for simulations of identical impacts at different (approximate) target resolutions, $N$, and maximum smoothing lengths, $h_{\rm max}$, as derived from the photic surface temperature maps shown in Figure \ref{fig:all_thermal_ps_3D}. Also listed are the masses of the target and impactor, $M_t$ and $M_i$, the target and impactor radii, $R_t$ and $R_i$, and the corresponding collision velocity, $v_c$, for each simulation. Non-exact values of $h_{\rm max}$, $\rho_{\min}$ and $v_c$ are given to 3 decimal places. $R_t$, $M_t$, $T_{\rm eff}$, $L$ and $\lambda_{\rm peak}$ are given to 3 significant figures.}
    \label{tab:all_sim_data}
\end{table*}

\section{Assessing Alternative Silicate Vapour Opacity Metrics}
\label{appendix:metric_comparison}
Two silicate-esque absorptivity assessment metrics were described in \cite{Gammie_2016} and applied to the study of a partially-vaporised post-impact disk. The models describe Rosseland mean opacities, $\kappa$, as calculated in \cite{badnell2005_OPproject} and \cite{park2013} for Lunar soil and H-chondrite compositions respectively. In \cite{badnell2005_OPproject}, $\kappa$ is calculated by the atomic absorptivity code AUTOSTRUCTURE, and the opacity calculations are valid across temperature and density ranges of $3 \times 10^{3} \ {\rm K} < T < 10^{7} \ {\rm K}$ and $10^{-12} \ \rm{kg \ m^{-3}} < \rho < 10 \ \rm{kg \ m^{-3}}$. Whilst this density range covers that of the photospheres found in this study, we must consider materials at lower temperatures than 3000K. Furthermore, these are atomic opacities only. \cite{Gammie_2016} also fit a power law equation for $\kappa$ to molecular absorption data from work presented in \cite{park2013}, allowing $\alpha_{\rm{vap}}$ to be estimated as:

\begin{equation}
    \label{eqn:park_opacity_eqn}
    \alpha_{\rm{vap}} = \kappa \ \rho_{\rm vap} \approx 55 \ \rho_{\rm vap}^{1.61} \ \bigg( \frac{T_{\rm vap}}{5000 \rm \, K} \bigg)^{1.2} \ \rm{m^2 \ kg^{-1}} \; ,
\end{equation}
for $T_{\rm vap}$ in units of K, and $\rho_{\rm vap}$ in units of $\rm{kg \ m^{-3}}$. This equation was fit to data spanning temperature and density ranges of $3 \times 10^{3} \ \rm{K} < T < 2 \times 10^{4} \ \rm{K}$ and $10^{-2} \ \rm{kg \ m^{-3}} < \rho < 10 \ \rm{kg \ m^{-3}}$. Unfortunately, as a power law fit to data, Equation \ref{eqn:park_opacity_eqn} is ill-suited to extrapolation to the T and $\rho$ ranges we will need to consider in this study.

A more sophisticated way of calculating the silicate vapour opacity would be to use molecular line lists, such as those
collated in databases like ExoMol \citep{exomol}, which have previously been used by \cite{nguyen2025_rockvapouropaque} to determine rock vapour opacity for lava worlds. ExoMol provides wavelength-dependent absorptivity and has line lists for $\rm{SiO_2}$ which could, as above, be used as a proxy for a silicate. The metric in \cite{kraus2012_vap} considers wavelength-averaged emission/absorption, and so for comparison, a simplified wavelength-averaged ExoMol absorptivity metric for molecular vapour can be given by:
\begin{equation}
    \label{eqn:exomol_vapour_absorption}
    \alpha_{\rm vap} = \frac{\bar{\sigma}(T_{\rm vap}) \cdot \rho_{\rm vap}}{M_a}
\end{equation}
where $M_a$ is mean molecular mass in kg per molecule, and $\bar{\sigma}(T_{ \rm vap})$ is wavelength-averaged absorbing cross-section as a function of vapour temperature, in units of $\rm{m^{2}}$ per molecule. The ExoMol linelist for $\rm{SiO_2}$ \citep{sio2_exomol_linelist} is calculated over a wavelength range of $\sim 1.7 - 100 \rm{\mu m}$ (near- to far-infrared). Unfortunately, the temperature range is limited to $1-3000$ K, which is too low in the context of post-giant impact bodies but, within this temperature range, we can broadly constrain (infrared) wavelength-averaged absorbing cross-section as ranging from $10^{-34}$ -- $10^{-22} \ \rm{m^{2}}$ per molecule for any $T_{\rm{vap}}$. 

As shown in Figure \ref{fig:absorptivity_metric_comparison}, \cite{kraus2012_vap} and \cite{Gammie_2016} demonstrate somewhat converse behaviour to Equation \ref{eqn:exomol_vapour_absorption} -- the transition between optically thick to thin regimes is temperature-dominated for the former, and density-dominated for the latter. Indeed, at very low densities, Equation \ref{eqn:park_opacity_eqn} shows very little variation with temperature (over the temperature range relevant for terrestrial giant impacts). Although Equation \ref{eqn:park_opacity_eqn} is an unreliable extrapolation of a power law fit, it agrees with the bounds imposed by Equation \ref{eqn:exomol_vapour_absorption}. The lower bound plotted for ExoMol is extremely conservative; we would expect the true ExoMol value to be several orders of magnitude larger than the plotted lower bound. At temperatures lower than $\sim3400$~K, absorptivity values calculated using \cite{kraus2012_vap} are either comparable or significantly smaller than the other two metrics. Our use of \cite{kraus2012_vap} compared to the two alternative metrics yields an approximate lower bound for opacity assessments.

\section{Volume Integration Approximation}
\label{appendix:volume}
In Subsection~\ref{subsect:data_processing:checks}, we present evidence that the introduction of massless pseudo-particles to evaluate material properties at arbitrary locations does not violate principles of mass and/or internal energy conservation. The results of this validation are presented in Figure~\ref{fig:mass_dist_comparison}: the mass and internal energies in different radial bins from the centre of mass of the post-impact structure is evaluated both by counting the masses of particles whose central locations are in that bin, and by integrating the interpolated mass and internal energy values within each radial bin. In this section, we describe the method used to assess the mass and energy in each radial volume slice, or ``shell'', denoted $i$.

We want to quantify the mass and internal energy that has been smoothed over each radial bin, assuming no artificial cap on smoothing length, $h_{\rm max}$. At an arbitrary location $\mathbf{x}$, we calculate density, $\rho_{\rm tot}$ via Equation~\ref{eqn:sph_rho} with a proxy smoothing length, and internal energy per volume, $\upsilon_{\rm tot}$, as follows:
\begin{equation}
    \upsilon_{\rm tot} = {u}_{\rm tot} \ \rho_{\rm tot}
\end{equation}
where specific internal energy, $u_{\rm tot}$ is calculated as:
\begin{equation}
    {u}_{\rm tot} = {u}_{\rm fe}({\rho}_{\rm fe}, {s}_{\rm fe}) \ \frac{m_{\rm fe}}{m_{\rm tot}}  + {u}_{\rm fors}({\rho}_{\rm fors}, {s}_{\rm fors}) \ \frac{m_{\rm fors}}{m_{\rm tot}} 
\end{equation}
Here, the constituent material parts of density, (${\rho}_{\rm fe}$ and ${\rho}_{\rm fors}$), and specific entropies of material parts (${s}_{\rm fe}$, ${s}_{\rm fors}$) are calculated as described in Subsection~\ref{subsect:data_processing:arbitrary}. Specific internal energies of material parts, ${u}_{\rm fe}$ and ${u}_{\rm fors}$, are calculated from the relevant EoS using the \texttt{A1\_Z\_rho\_Y} function in WoMa.

\begin{figure}
    \centering
    \includegraphics[width=\linewidth]{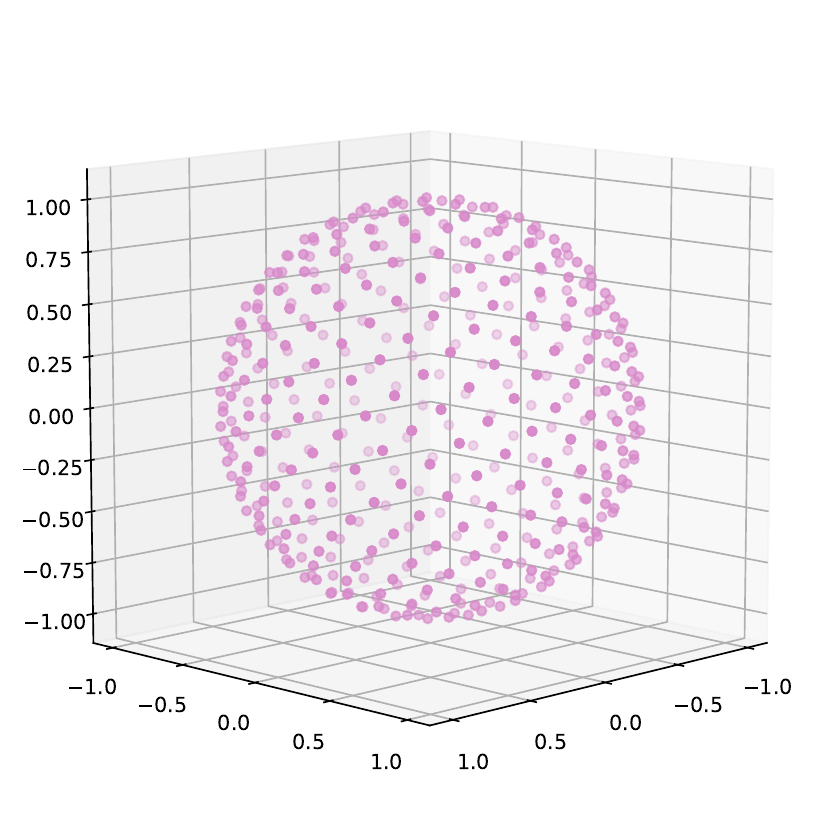}
    \caption{A Fibonacci lattice of $\sim$~250 points arranged on the surface of a unit sphere using a Fibonacci algorithm, such that they are suitably equally-spaced. This allows us to more accurately approximate a volume integral to verify the conservation of both mass and internal energy after the application of our post-processing algorithm.}
    \label{fig:fib_algo}
\end{figure}

Calculating masses per volume and internal energies per volume for a range of sample points in each radial bin allows us to estimate a volume integral of $\rho_{\rm tot}$ and $\upsilon_{\rm tot}$ over each radial shell. This means we can approximate the total mass and total energy in each radial bin, and thus the entire information region of our impact structure. However, due to the computational expense of our pseudo-particle algorithm, especially in regions of particle sparsity, we are limited to how many sample points we can choose in each radial shell. We also want to ensure that the limited selection of sample points in each radial shell is somewhat spatially equidistributed to avoid oversampling in any given direction (e.g., sampling more points at the poles of the post-impact object along its rotation axis rather than in the plane of impact). To do this, we use a Fibonacci algorithm \citep{Gonzalez_2009_fiblattice} to generate an approximately equally-distributed collection of points on the surface of the unit sphere, a so-called Fibonacci lattice, as demonstrated in Figure~\ref{fig:fib_algo}. The longitudes, $\theta_j$, and latitudes, $\varphi_j$, of a Fibonacci lattice with $2P+1$ points are generated as follows:
\begin{align}
    \theta_j &= 2\pi\phi^{-1} \ {\rm mod}(j,\phi)  \\
    \varphi_j &= {\rm arcsin}\bigg(\frac{2j}{2P + 1}\bigg) + \frac{\pi}{2}
\end{align}
for $j \in \mathbb{Z}$ such that $-P \leq j \leq P$, where $\phi = \frac{1 + \sqrt{5}}{2}$ is the golden ratio. These angles are in radians such that $\theta_j \in [0,2\pi]$ and $\varphi_j \in [0,\pi]$, so that these spherical coordinate values can easily be converted into Cartesians \citep[formulae are adapted from][]{Gonzalez_2009_fiblattice}. Furthermore, we can project the Fibonacci lattice onto larger spheres by scaling the radius of each point.

Now that we can select a chosen resolution of $n = 2P+1$ approximately equidistributed sample points in each radial bin, we can approximate volume integrals of $\rho_{\rm tot}$ and $\upsilon_{\rm tot}$ over each shell $i$ to calculate $m_i$ and $U_i$ in each shell:
\begin{align}
    m_i &= \frac{4 \pi}{3} (R_{i+1}^3 - R_{i}^3) \frac{1}{n} {\sum}_{k=1}^{n} \rho_k(\mathbf{x}_k) \; , \\
    U_i &= \frac{4 \pi}{3} (R_{i+1}^3 - R_{i}^3) \frac{1}{n} {\sum}_{k=1}^{n} \upsilon_k(\mathbf{x}_k) \; ,
\end{align}
where $R_i$ and $R_{i+1}$ are the radii of the inner and outer edge of each shell, and $\{ \mathbf{x}_k \ |\ 1 \leq k \leq n \}$ is a Fibonacci lattice projected onto a sphere of radius $\frac{1}{2}(R_{i} + R_{i+1})$. These approximations of radially binned interpolated mass, $m_i$, and interpolated internal energy, $U_i$, are plotted in Figure~\ref{fig:mass_dist_comparison} alongside radially binned particle values. Finally, we calculate $M_{\rm tot}$ and $U_{\rm tot}$ across all bins simply by summing over all bins:
\begin{align}
    M_{\rm tot} &= \sum_{i} \ m_i \; , \\
    U_{\rm tot} &= \sum_{i} \ U_i \; .
\end{align}
$M_{\rm tot }$ and $U_{\rm tot }$ are checked against the mass and internal energy totals calculated from raw particle data before processing, as discussed in Subsection~\ref{subsect:data_processing:checks}.

\section{Condensate Fractions by Mass and Volume}
In this section we present plots of condensate fraction by volume (see Equation~\ref{eqn:cond_terms_vol}) and condensate fraction by mass. The latter can be interpolated (ideally from post-processed data) using an analogous expression to vapour mass fraction (see Equation~\ref{eqn:vap_mass_frac}). In this context, since our method assumes material to either be liquid, vapour, or a mix of those two phases, we can alternatively use $1 -$ Equation~\ref{eqn:vap_mass_frac} to find condensate mass fraction.

\begin{figure*}
    \centering
    \includegraphics[width=\linewidth]{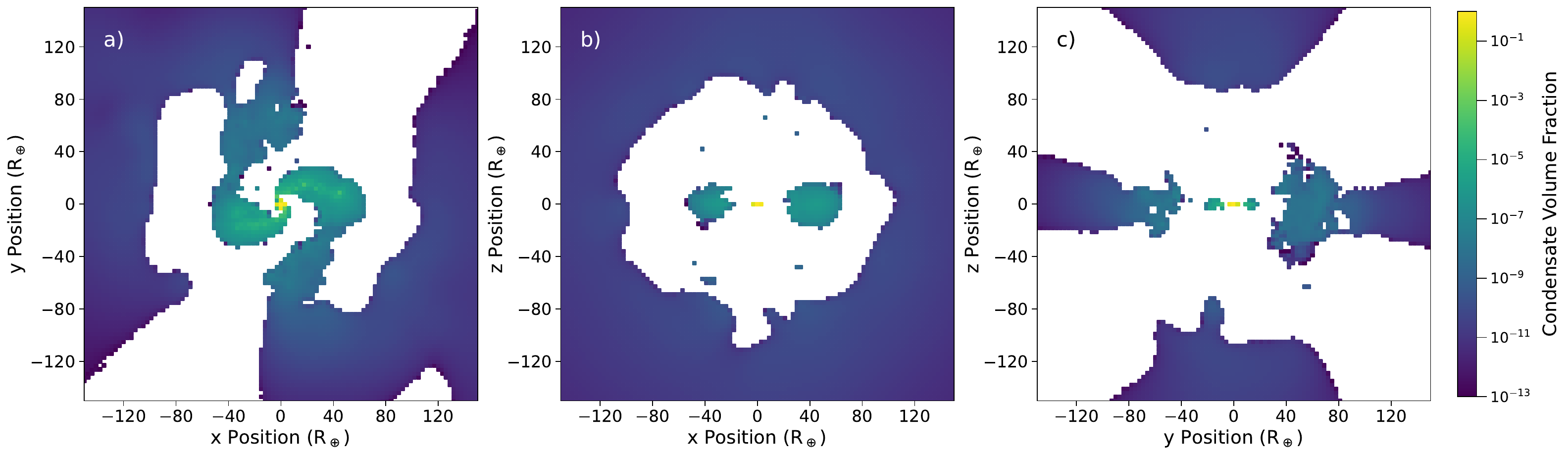}
    \caption{Condensate volume fraction through the planes a) $z = 0$, b) $y=0$ and c) $x=0$, calculated using Equation~\ref{eqn:vap_mass_frac} at 24\,h into the terrestrial giant impact example used in this paper, with $N \sim 10^{6.5}$ and $h_{\rm max} = 1R_\oplus$. This volume fraction is log-scaled, highlighting that other than very centrally, vapour is the major phase by volume throughout this structure.}
    \label{fig:cond_frac_vol_example}
\end{figure*}

We note that the absorptivity metric used in this study, Equation~\ref{eqn:photon_prob}, relies on the assumption that it is being evaluated at locations where volume is dominated by vapour. In Figure \ref{fig:cond_frac_vol_example}, we show 2D slices through both the plane of impact ($z=0$), and two perpendicular planes partially defined by the rotation axis (defined by $x=0$ and $y=0$), throughout which condensate fraction by volume has been calculated. These plots suggest that at most locations in the impact structures resulting from this impact example, other than near the origin (the centre of mass), condensate fraction by volume is incredibly close to zero, which justifies the assumption that vapour is the major phase by volume in the regions over which our metric is applied.

\label{appendix:photic_surface_condensate}

\begin{figure*}
    \centering
    \includegraphics[width=\textwidth]{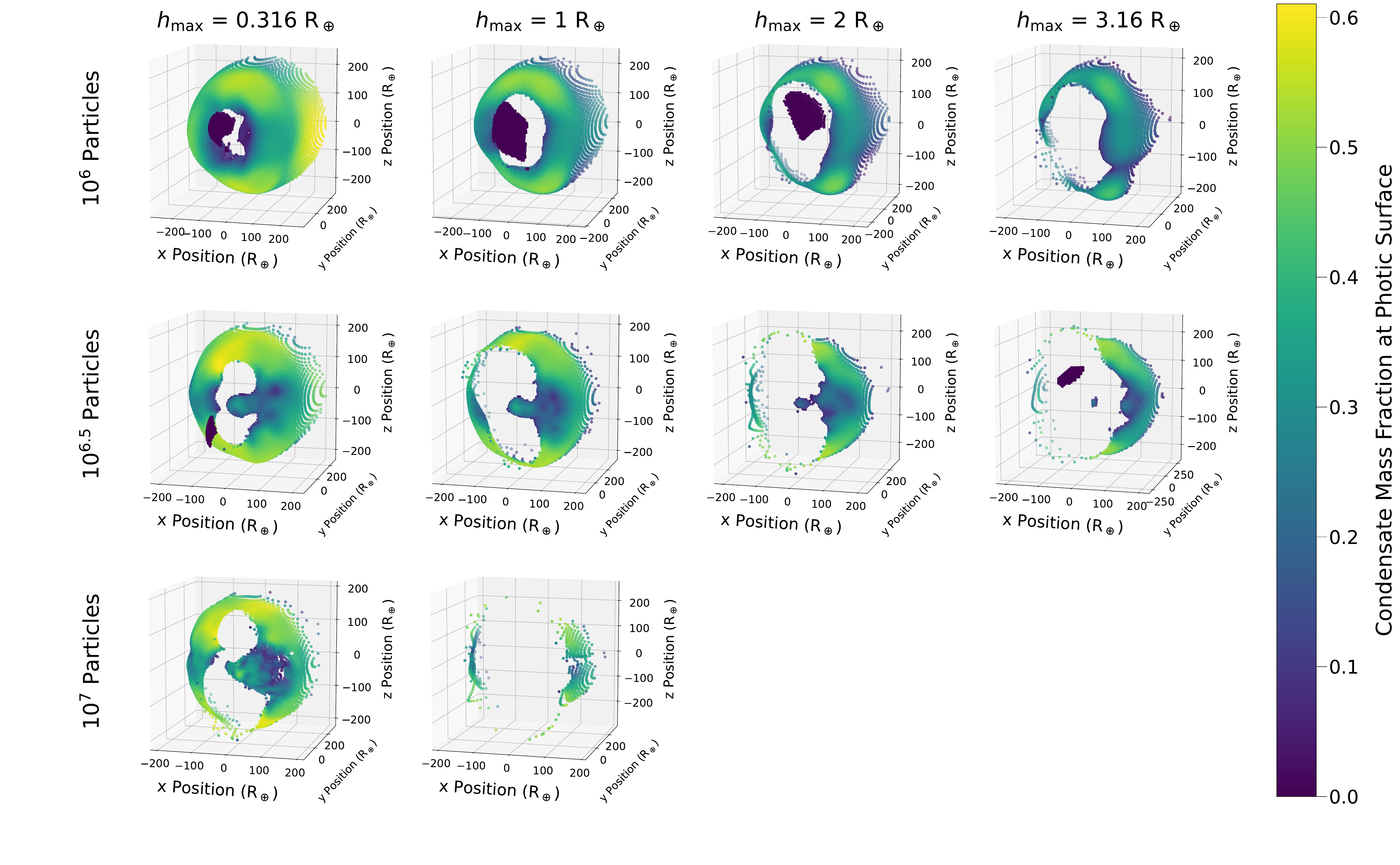}
    \caption{A grid of extracted photic surfaces resulting from the same giant impact scenario between two terrestrial bodies plotted in three dimensions, at a simulation time of 24\,h. Each row corresponds to the same approximate target resolution, each column to the same maximum smoothing length, $h_{\rm max}$. The surfaces $\tau$ = 1 are colour-coded by condensate mass fraction. When no colour is used, this either indicates that the line of sight is optically thin, or that a line of sight is optically thick but condensate mass fraction is 0 at the photic surface. In the case that the latter is true, no colour is used in order to emphasise a complete absence of condensate.}
    \label{fig:cond_frac_all_ps}
\end{figure*}

Furthermore, complementary to the results presented in Figure~\ref{fig:all_thermal_ps_3D}, Figure~\ref{fig:cond_frac_all_ps} shows the grid of extracted photic surfaces resulting from our terrestrial giant impact scenario, this time demonstrating the condensate mass fraction, as opposed to the temperature, at each location. This figure demonstrates that the heated ``plumes'' seen in Figure~\ref{fig:all_thermal_ps_3D} are almost entirely composed of vapour, since the condensate mass fraction is either zero or close to zero in these locations. This implies that in these regions, opacity is dominated by vaporised impact material, rather than condensate.

%%%%%%%%%%%%%%%%%%%%%%%%%%%%%%%%%%%%%%%%%%%%%%%%%%

%% For this sample we use BibTeX plus aasjournalv7.bst to generate the

\bibliography{paper1_lucy}{}
\bibliographystyle{aasjournalv7.1}



\end{document}